\documentclass[aps,pra,superscriptaddress,reprint]{revtex4-1}
\usepackage{amsmath, amssymb}    
\usepackage{graphicx}   
\graphicspath{ {./img/} }
\usepackage{verbatim}   
\usepackage{subfigure}  
\usepackage{hyperref}   
\usepackage{pstricks}
\usepackage{psfrag}
\usepackage{datatool, tikz, pgf}
\usepackage{overpic}

\newcommand{\psihat}{\hat{\psi}}
\newcommand{\phihat}{\hat{\phi}}
\newcommand{\rt}{(\mathbf{r},t)}
\newcommand{\Lloc}{L_{\text{loc}}}
\newcommand{\ket}[1]{\ensuremath{|#1\rangle}}
\newcommand{\bra}[1]{\ensuremath{\langle #1|}}
\newcommand{\elemm}[3]{\ensuremath{\bra{#1}#2\ket{#3}}}
\newcommand{\braket}[2]{\ensuremath{\langle #1|#2\rangle}}

\begin{document}
\title{Breakdown of Anderson localization in the transport of Bose-Einstein condensates through one-dimensional disordered potentials}
\author{Julien Dujardin}
\affiliation{D\'{e}partement de Physique, University of Liege, 4000 Li\`ege, Belgium}
\author{Thomas Engl}
\affiliation{Institut f\"ur Theoretische Physik, Universit\"at Regensburg, D-93040 Regensburg, Germany}
\author{Peter Schlagheck}
\affiliation{D\'{e}partement de Physique, University of Liege, 4000 Li\`ege, Belgium}

\begin{abstract}
We study the transport of an interacting Bose--Einstein condensate through a 1D correlated disorder potential. We use for this purpose the truncated Wigner method, which is, as we show, corresponding to the diagonal approximation of a semiclassical van Vleck--Gutzwiller representation of this many-body transport process. We also argue that semiclassical corrections beyond this diagonal approximation are vanishing under disorder average, thus confirming the validity of the truncated Wigner method in this context. Numerical calculations show that, while for weak atom-atom interaction strength Anderson localization is preserved with a slight modification of the localization length, for larger interaction strenghts a crossover to a delocalized regime exists due to inelastic scattering. In this case, the transport is fully incoherent.
\end{abstract}
\maketitle

\section{Introduction}
Localization phenomena in disordered systems highlight the fundamental role of interference of wave propagation \cite{Akkermans1995}. A particular example of this is Anderson localization \cite{Anderson1958PR} (AL) describing the metal-insulator quantum phase transition in a 3D medium containing random impurities. The waves are coherently scattered over the impurities several times and yield to waves with  an exponential profile resulting in a suppression of transport. Anderson localization has been experimentally observed with light in diffusive media \cite{Wiersma1997N} and in photonic crystals \cite{Schwartz2007N}, with microwaves \cite{Chabanov2000N}, and with sound waves \cite{Hu2008NP}.

The recent progresses in the field of ultracold atoms has opened the possibility to experimentally study AL with a Bose-Einstein condensate (BEC), both in momentum space \cite{Chabe2008PRL} with a $\delta$-kicked BEC \cite{Moore1995PRL} as well as in real space with the direct observation of exponential tails of the density profiles \cite{Billy2008N,Roati2008N}. The use of a BEC to study AL is very appealing since it is possible to precisely tune the different properties of the BEC such as, for instance, the two-body interaction strength between bosonic atoms via Feshbach resonances \cite{Fano1961PR,Feshbach1962AoP,Stwalley1976PRL}. Research in the field of localization of bosonic atoms is motivated by the direct analogy with the electronic counterpart \cite{Brantut2012S,Bruderer2012PRA,Kristinsdottir2013PRL,Brantut2013S} and the perspective to realize bosonic atomtronic devices \cite{Micheli2004PRL,Daley2005PRA,Seaman2007PRA,Pepino2009PRL} such as an atomic transistor. In this, case the interplay between localization effects due to impurities and two-body interaction can play a major role in transport properties.

An accurate and numerically efficient theoretical description of transport process in such an open system faces the challenge of dealing with a many-body system with a potentially high number of atoms. The size of the associated Hilbert space renders techniques such as exact diagonalization or Density Matrix Renormalization Group (DMRG) methods inefficient or even practically impossible with the computational resources that are nowadays available. On the other hand, in the mean-field limit, corresponding to a high number of atoms and a weak two-body interaction, it is possible to describe the dynamics of the BEC in an approximate manner using the nonlinear Gross--Pitaevskii (GP) equation. This limit has been extensively studied in the context of bosonic transport \cite{Leboeuf2001PRA,Carusotto2001PRA,Paul2005PRA,Paul2005PRL,Paul2007PRA,Ernst2010PRA}. Furthermore, it has been shown that, even for weak interactions between atoms, inelastic scattering processes are not negligible \cite{Geiger2012PRL,Geiger2013NJoP} thus potentially invalidating results obtained by the GP mean-field description. Several methods exists to take into account effects beyond the mean-field description. In particular a Bogoliubov description \cite{Bogoliubov1947,Huang1992PRL,Giorgini1994PRB,Oosten2001PRA,Kobayashi2002PRB,Kuhn2007NJoP,Gaul2011PRA,Mueller2012NJoP,Gaul2013TEPJST,Gaul2014APB} seems very promising and offers the advantage of allowing for analytical solutions. 

As was done in Refs.~\cite{Paul2005PRA,Ernst2010PRA} we study, in this paper, the effect of atom-atom interactions on AL in the context of an atom laser \cite{Guerin2006PRL,Couvert2008EEL,Vermersch2011PRA,Bolpasi2014NJP} which consists of a reservoir of atoms that are outcoupled into a waveguide. The atoms eventually encounter the disordered region within this waveguide. We describe the system with the truncated Wigner (tW) method  \cite{Wigner1931,Wigner1932PR,Moyal1949PCPS,GarZol,Steel1998PRA,Sinatra2002JPBAMOP,Polkovnikov2003PRA} that essentially consists in representing the many-body bosonic quantum fields in terms of classical fields. This method has been successfully used to study several dynamical processes in closed systems, see \textit{e.g.} Refs.~\cite{Isella2006PRA,Schmidt2012NJoP}. The steps towards a numerically efficient formulation of the truncated Wigner method for a one-dimensional open systems have been achieved in Refs.~\cite{Dujardin2014APB,Dujardin2015PRA}. Moreover, the tW method has been shown to correctly model inelastic scattering \cite{Dujardin2015PRA} and to yield consistent results in the continuous limit \cite{Dujardin2015AdP}.

The main object of our study is the breakdown of AL in correlated disorder potentials due to the presence of atom-atom interaction, which is to be investigated with the truncated Wigner method. To this end, we first describe, in Sec.~\ref{sec:guidedatomlaser}, the one-dimensional transport scenario we are interested in and then, in Sec.~\ref{subsec:TW}, describe how the truncated Wigner method can be implemented. In Sec.~\ref{subsec:semiclassicvalidityoftruncatedwigner}, using a semiclassical approach \cite{Engl2014PRL}, we compute the leading corrections to the so-called diagonal approximation, the latter being equivalent to the truncated Wigner method \cite{Dujardin2015AdP}. These leading corrections are shown to be negligibly small when performing a disorder average. We finally compare numerically, in Sec.~\ref{sec:results}, the transmission and the density profiles across Gaussian-correlated disorder potentials provided by the mean-field GP description and the truncated Wigner method.

\section{Guided atom lasers}
\label{sec:guidedatomlaser}
\subsection{Theoretical description}
\begin{figure}[t]
  \centering
  \includegraphics[width=0.95\linewidth]{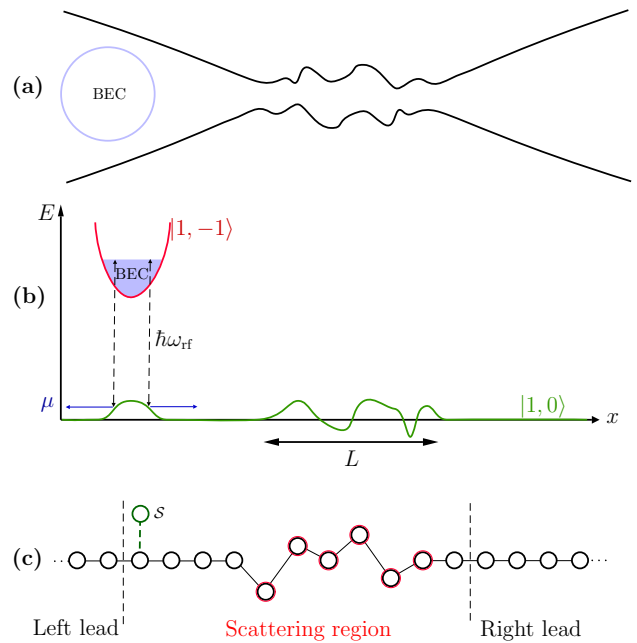}
  \caption{(color online) (a) Typical atom laser configuration inspired from Ref.~\cite{Guerin2006PRL}, consisting of a BEC within a magnetic trap of atoms and an optical waveguide generated by a spatially focused light field. (b) The atoms in the reservoir are outcoupled from the trapped state $\ket{F, m_F} = \ket{1,-1}$ to the untrapped state $\ket{1,0}$ thanks to a radio-frequency field of energy $\hbar\omega_{\text{rf}}$. In the untrapped state, atoms are confined in the waveguide rendering the propagation quasi one-dimensional in the longitudinal $x$-direction. A disorder potential can then be engineered to study transport properties across it. (c) The corresponding theoretical description involves a spatial grid that is described in Sec.~\ref{subsec:discspace}. We suppose, in addition, that the potential has a finite support of length $L$ and that atoms are interacting only within this support.}
  \label{fig:atlaser}
\end{figure}

Let us consider a typical atom laser experiment such as the one reported in Ref.~\cite{Guerin2006PRL}, the working principle of which is depicted in Fig.~\ref{fig:atlaser}. The condensed atoms of $^{87}$Rb are trapped in a magnetic trap and constitute the \emph{reservoir} containing a high number $N$ of atoms that are ideally in a perfect condensate maintained at zero temperature. Theses atoms are spin-polarized in the trapped state $\ket{F, m_F} = \ket{1,-1}$ with energy $E_1$. To release the atoms, \textit{i.e.} transfer them to an untrapped state $\ket{1,0}$ with energy $E_0$, a radio-frequency field consisting in a nearly resonant electromagnetic wave at energy $\hbar\omega_{\text{rf}}$ ($\sim 2\pi \hbar\cdot$ GHz) is used to couple the state $\ket{1,-1}$ to the state $\ket{1,0}$. These atoms are insensitive to the external magnetic field and propagate in the waveguide. The chemical potential is given by $\mu=E_1-E_0-\hbar\omega_{\text{rf}}$ and corresponds to the kinetic energy per atom. The propagation of atoms can thus be considered as a free quasi one-dimensional motion. It is then possible to engineer a correlated disorder potential (see \textit{e.g.} Ref.~\cite{Clement2006NJoP}) with a finite extent $L$ and study the suppression of transport.

In order to represent the atoms in the atom laser system, we define, in second quantizaztion, the field operator of the atoms in the trap $\phihat_{m_{F=-1}}\rt$ corresponding to the atoms in the hyperfine state $\ket{1,-1}$, and the field operator of the atoms in the waveguide $\psihat_{m_{F=0}}\rt$ corresponding to the atoms in the hyperfine state $\ket{1,0}$ as
\begin{subequations}
  \begin{align}
    \psihat_{m_{F=0}}\rt &= \psihat\rt e^{-i\mu t/\hbar},\\
    \phihat_{m_{F=-1}}\rt &= \phihat\rt e^{-i(\mu +\hbar\omega_{\textrm{rf}})t/\hbar}.
  \end{align}
\end{subequations}
The corresponding evolution equation for the field operators in the Heisenberg representation can be written as
\begin{subequations}
  \label{eq:evoleq}
  \begin{align}
    i\hbar\frac{\partial}{\partial t} \psihat\rt &= \left[-\frac{\hbar^2}{2m}\frac{\partial^2}{\partial \mathbf{r}^2} -\mu + V_
{\textrm{opt}}(\mathbf{r}) \right]\psihat\rt \nonumber\\
                                                 &+ \mathcal{U}\psihat^\dagger\rt\psihat\rt\psihat\rt\nonumber\\
                                                 &+ D(t)\,\phihat\rt, \\
    \label{eq:evoleqtwo}
    i\hbar\frac{\partial}{\partial t} \phihat\rt &= \left[-\frac{\hbar^2}{2m}\frac{\partial^2}{\partial \mathbf{r}^2} + V_{\textrm{trap}}(\mathbf{r}) \right]\phihat\rt \nonumber\\
                                                 &+ D^*(t)\,\psihat\rt,
  \end{align}
\end{subequations}
where $V_{\textrm{opt}}$ is the optical potential including the waveguide and $V_{\textrm{trap}}$ is the magnetic trapping potential. We consider a slowly varying transverse confinement frequency $\omega_\perp(x)$ of the waveguide in such a way that the spatial variations of this potential can not induce excitation of transverse modes. We take into account the interactions between the atoms by the mean of a low-energy contact pseudopotential \cite{Pethick2008}. The interaction strength between the atoms in state $\ket{1,0}$ is $\mathcal{U}=4\pi\hbar^2a_S/m$, where $a_S$ is the $s$-wave scattering length. The coupling strength between the reservoir and the guide is defined as $D(t)$ and is adiabatically ramped from zero to a constant value $D$.

Assuming that the atoms in the reservoir are all in the ground state, we can decompose the associated field operator as  $\phihat\rt=\phi_0(\mathbf{r})\hat{\psi}_S(t)$. During the coupling process, we suppose that only the transverse ground mode $\chi_0(y,z)$ is populated. We can therefore decompose the field operator associated with the waveguide as $\psihat\rt~=~\chi_0(y,z)\psihat(x,t)$. Finally we neglect both the (intra-species) interaction between ($m_F=-1$) atoms in the reservoir and the (inter-species) interaction between ($m_F=-1$) atoms in the reservoir and ($m_F=0$) atoms in the waveguide, assuming that the latter do not play any role in the vicinity of the disordered system. These assumptions allow us to simplify Eqs.~\eqref{eq:evoleq} yielding
\begin{subequations}
  \label{eq:evoleqsimplified}
  \begin{align}
    i\hbar\frac{\partial}{\partial t}\psihat(x,t) &= 
    \mathcal{H}_0\psihat(x,t) + g(x)\psihat^\dagger(x,t)\psihat(x,t)\psihat(x,t) 
\nonumber \\ & + K(x,t)\hat{\psi}_S(t),\\
    i\hbar\frac{\partial}{\partial t} \hat{\psi}_S(t)  &= \int dx \, K^*(x,t)\hat{\psi}(x,t),
  \end{align}
\end{subequations}
where we define
\begin{subequations}
  \begin{align}
    \mathcal{H}_0 &= \mathcal{H}_k-\mu+V(x) ,\\
    \mathcal{H}_k &= -\frac{\hbar^2}{2m}\frac{\partial^2}{\partial x^2}, \\
    g(x)          &= 2\hbar\omega_\perp(x)a_S,\\
    K(x,t)        &= D(t) \iint dy\,dz\, \chi_0^*(y,z)\,\phi_0(\mathbf{r}).
  \end{align}
\end{subequations}

\subsection{Discretization of space}
\label{subsec:discspace}
Following the procedure described in Ref.~\cite{Dujardin2015AdP}, we discretize space by constructing a grid along the one dimensional guide by introducing an energy cutoff in momentum space at $p=\pm\pi\hbar/\delta$. We modify $\mathcal{H}_k$ accordingly by
\begin{equation}
  \mathcal{H}_k^\delta  = E_\delta \left[1- \cos(\delta \hat{p}/\hbar)\right],
\end{equation}
with $\hat{p}= -i\hbar\partial_x$ and $E_\delta=\hbar^2/m\delta^2$. The eigenstates of the modified Hamiltonian  $\tilde{\phi}_k(x)=\exp(ikx)/\sqrt{2\pi}$ are identical to the ones of $\mathcal{H}_k$ and the associated eigenvalues now read
\begin{equation}
  E_k = E_\delta[1-\cos(k\delta)],
\end{equation}
where $k\in[-\pi/\delta,\pi/\delta]$. In the continuous limit, \textit{i.e.} for $k\delta \to 0$, we have $\mathcal{H}_0^\delta \to \mathcal{H}_0$ and the eigenvalues tend to the continuous ones: $E_k = \hbar^2 k^2 / 2m$. 

Defining an effective Wannier basis composed of spatially localized functions
\begin{equation}
  \phi_l(x) = \sqrt{\frac{\delta}{2\pi}}\int_{-\pi/\delta}^{\pi/\delta}\tilde{\phi}_k(x) e^{ilk\delta}\,dk
\end{equation}
for $l\in\mathbb{Z}$, and the corresponding bosonic annihilation $\psihat_l$ and creation $\psihat^\dagger_l$ operators, we have
\begin{equation}
  \label{eq:wannierdevelop}
  \psihat(x)=\sum_{l=-\infty}^\infty \phi_l(x)\psihat_l.
\end{equation}
In addition, we suppose that the source is located at position $x_S$ corresponding to the site $l_S$. As a consequence, we can define a coupling strength $\kappa(t)$ on site $l_S$ as
\begin{equation}
  \label{eq:wanniersource}
  K(x,t) = \kappa(t) \phi_{l_S}(x).
\end{equation}

Inserting Eqs.~(\eqref{eq:wannierdevelop} and \eqref{eq:wanniersource}) in the evolution equations \eqref{eq:evoleqsimplified}, we obtain, in the limit $\delta \to 0$,
\begin{subequations}
  \label{eq:inffieldopevol}
  \begin{align}
    i\hbar \frac{\partial \psihat_l(t)}{\partial t} &= (E_\delta + V_l - \mu)\psihat_l(t)- \frac{E_\delta}{2}\left[\psihat_{l-1}(t) + \psihat_{l+1}(t)\right] \nonumber\\
                                                    &+ g_l \psihat_l^\dagger(t)\psihat_l(t)\psihat_l(t) + \kappa(t)\delta_{l,l_S}\psihat_{S}\\
    i\hbar \frac{\partial \psihat_S(t)}{\partial t} &= \kappa^*(t)\psihat_{l_S}(t),
  \end{align}
\end{subequations}
with $V_l = V\left(l\delta\right)$,  $g_l = g\left(l\delta\right)/\delta$, $E_\delta=\hbar^2/(m\delta^2)$ where the coupling strength is well approximated by $\kappa(t)\simeq K(x_S,t)/\sqrt{\delta}$. A visual representation of this discretization scheme is provided in Fig.~\ref{fig:atlaser}(c).

\section{Truncated Wigner method}
\label{subsec:TW}
The truncated Wigner method \cite{Wigner1931,Wigner1932PR,Moyal1949PCPS,GarZol,Steel1998PRA,Sinatra2002JPBAMOP,Polkovnikov2003PRA} has been recently adapted to deal with many-body bosonic scattering processes in the context of a guided atom laser \cite{Dujardin2015PRA,Dujardin2015AdP}. Describing the latter with the evolution equations defined by Eqs.~\eqref{eq:inffieldopevol}, applying the tW method effectively amounts to solving the equations \cite{Dujardin2015PRA}
\begin{subequations}
  \label{eq:StochAll}
  \begin{align}
    i\hbar\frac{\partial \psi_l}{\partial t} &= (E_\delta + V_l-\mu)\psi_l - \frac{E_\delta}{2}\left(\psi_{l+1} + \psi_{l-1}\right)  \nonumber\\
                                             &+ g_l(|\psi_l|^2-1) \psi_l + \kappa(t)\psi_S\delta_{l,l_S}, \\
   \label{eq:StochSource}
    i\hbar\frac{\partial \psi_S}{\partial t} &=  \kappa^*(t)\psi_{l_S},
  \end{align}
\end{subequations}
where $\psi_l$ and $\psi_S$ are the classical fields associated with site $l$ and the source, respectively. Denoting by $\mathcal{A}=\{S,0,\pm1,\pm2,\cdots\}$ the ensemble of sites of the system, we can express the corresponding classical Hamiltonian $H_{\textrm{cl}}(\boldsymbol{\psi},\boldsymbol{\psi}^*)$, where $\boldsymbol{\psi}=(\psi_\alpha)_{\alpha\in \mathcal{A}}$, as
\begin{eqnarray}
  \label{eq:classicalhamiltonian}
  H_{\textrm{cl}}(\boldsymbol{\psi},\boldsymbol{\psi}^*) &=& \sum_{l\in \mathbb{Z}}\Big[ (E_\delta + V_l-\mu)|\psi_l|^2 + \frac{g_l}{2} |\psi_l|^2(|\psi_l|^2-2) \nonumber \\
               & & - \frac{E_\delta}{2}\left(\psi^*_{l+1}\psi_l + \psi^*_l\psi_{l+1}\right) \Big] \nonumber \\
               & & + \kappa^*(t)\psi_S^*\psi_{l_S} + \kappa(t)\psi^*_{l_S}\psi_S.
\end{eqnarray}

The classical fields $\psi_l$ and $\psi_S$ correctly sample, in the framework of the tW method, the initial quantum state. More specifically, for the empty waveguide, the initial field amplitudes at $t=t_0$, read
\begin{equation}
  \label{eq:InitCond}
  \psi_l(t_0) = \left(A_l + i B_l\right)/2,
\end{equation}
with $A_l$ and $B_l$ real, independent Gaussian random variables with unit variance and zero mean, \textit{i.e.} for each $l,l' \in \mathbb{Z}$ we have
\begin{subequations}
  \label{eq:initavg}
  \begin{align}
    \overline{A_l}      &=  \overline{B_l} = 0,\\
    \overline{A_{l'}A_l} &=  \overline{B_{l'}B_l} = \delta_{l,l'},	\\
    \overline{A_{l'}B_l} &= 0,
  \end{align}
\end{subequations}
where the overline denotes the average over the random variables. Since $\overline{|\psi_l|^2}=1/2$, this sampling effectively amounts to having half a (pseudo) particle on each site of the numerical grid from the classical point of view. The source of atoms is a BEC with a high number $N$ of atoms and can therefore be represented as a coherent state $\ket{\psi_S^0}$. As $N$ is very large, the relative uncertainty of both the amplitude $|\psi_S^0| = \sqrt{N}$ and the associated phase of the source are negligibly small. We therefore treat the source term completely classically and set $\psi_S^0~=~\sqrt{N}$ without loss of generality. Supposing, in addition, that the coupling $\kappa(t)$ tends to zero such that $N|\kappa(t)|^2$ remains finite, we can safely neglect the depletion of the source or any back-action of the waveguide to the source. In this context, the evolution of the source can be neglected and we can focus on the evolution of the field in the waveguide only. The equation of evolution finally reads
\begin{eqnarray}
  \label{eq:StochFinal}
  i\hbar\frac{\partial \psi_l}{\partial t} &=& (E_\delta + V_l-\mu)\psi_l -\frac{E_\delta}{2}\left(\psi_{l+1} + \psi_{l-1}\right) \nonumber\\
                                           & &+ g_l(|\psi_l|^2 -1) \psi_l + \kappa(t)\sqrt{N}\delta_{l,l_S}.
\end{eqnarray}

The observables such as the density and the current are then computed through the statistical average over all classical initial states. This average will be denoted by an overline as in Eqs.~\eqref{eq:initavg}. For instance, the total density $\rho(x=l\delta,t)$ and current $j(x=l\delta,t)$ at position $x=l\delta$ are computed through 
\begin{eqnarray}
  \rho(l\delta,t) &=& \frac{1}{\delta}\left(\overline{|\psi_l(t)|^2} - 1/2\right), \\
  j(l\delta,t) &=& \frac{i\hbar}{2m\delta^2}\left( \overline{\psi^*_{l+1}(t)\psi_l(t) - \psi^*_l(t)\psi_{l+1}(t)} \right).
\end{eqnarray}
The half-particle that is added through the sampling of the initial state is subtracted after the time propagation. In the tW framework, it is possible to decompose the total current and the total density into coherent and incoherent contributions. The first contribution comes from the coherent parts of the density $\rho^{\text{coh}}(x=l\delta,t)$ and the current $j^{\text{coh}}(x=l\delta,t)$ at position $x=l\delta$, which represent the condensed atoms. They can be written as
\begin{eqnarray}
  \rho^{\text{coh}}(l\delta,t) &=& \left|\overline{\psi_l(t)}\right|^2/\delta,\\
  j^{\text{coh}}(l\delta,t) &=& \frac{i\hbar}{2m\delta^2}\left(   \overline{\psi^*_{l+1}(t)}\,\,\overline{\psi_l(t)} - \overline{\psi^*_l(t)}\,\,\overline{\psi_{l+1}(t)}\right).
\end{eqnarray}
The incoherent parts of the density $\rho^{\text{incoh}}(x=l\delta,t)$ and current $j^{\text{incoh}}(x=l\delta,t)$ at position $x=l\delta$ are defined as the difference between the total part and the coherent part. They read
\begin{eqnarray}
  \rho^{\text{incoh}}(l\delta,t) &=& n(l\delta,t) - n^{\text{coh}}(l\delta,t),\\
  j^{\text{incoh}}(l\delta,t) &=& j(l\delta,t) - j^{\text{coh}}(l\delta,t).
\end{eqnarray}

The total transmission is then naturally defined as the ratio of the total current in the downstream region and the stationary current $j^{\varnothing}$ for an homogeneous waveguide with a vanishing atom-atom interaction. It reads 
\begin{equation}
  \label{eq:totaltransmission}
  T = \lim_{t\to\infty} j(t)/j^{\varnothing}.
\end{equation}
where we dropped the spatial dependence of the current. The stationary current $j^{\varnothing}$ is given by
\begin{equation}
  \label{eq:freej}
  j^{\varnothing} = \frac{\delta}{\hbar}\frac{N|\kappa|^2}{\sqrt{\mu(2E_\delta-\mu)}}.
\end{equation}
In the same way, one can define the coherent part of the transmission and the incoherent part of the transmission as
\begin{subequations}
 \begin{align}
   T^{\text{coh}}   &= \lim_{t\to\infty} j^{\text{coh}}(t)/j^{\varnothing}, \\
   T^{\text{incoh}}  &= \lim_{t\to\infty} j^{\text{incoh}}(t)/j^{\varnothing}.
 \end{align}
\end{subequations}
Finally, we can also evaluate the stationary density $\rho^\varnothing$ in the case of an homogeneous waveguide with vanishing atom-atom interactions as
\begin{equation}
  \rho^\varnothing = \frac{1}{\delta} \frac{N|\kappa|^2}{\mu(2E_\delta-\mu)}.
\end{equation}

For the rest of the paper, we will consider that the on-site potential $V_l$ and the contact interaction strength $g_l$ are nonvanishing only in a finite region of space. The latter region will be called the \emph{scattering region} and the region on the left- and right-hand side of it will be named \emph{left} and \emph{right leads} in close analogy with electronic mesoscopic physics [see Fig.~\ref{fig:atlaser}(c)]. An efficient way to numerically model the decay of atom to the leads is provided by the method of Smooth Exterior Complex Scaling \cite{Dujardin2014APB}. This method transforms the infinite scattering system in a finite open system. The initial quantum fluctuations of the leads take the form of a time-dependent quantum noise that enters into the scattering region by the first and the last site \cite{Dujardin2015PRA}. This allows one to efficiently simulate scattering processes of guided matter waves using the tW method \cite{Dujardin2015PRA}.

\section{Validity of the truncated Wigner method}
\label{subsec:semiclassicvalidityoftruncatedwigner}

As was shown in Ref.~\cite{Dujardin2015AdP}, the tW method corresponds to the diagonal approximation \cite{da} in the framework of the semiclassical van Vleck-Gutzwiller description \cite{Gutzwiller} of the bosonic many-body system. This insight offers a novel framework for assessing the validity of the tW approach. We have to require, on the one hand, that the actions of the trajectories involved in the van Vleck-Gutzwiller approach are large compared to the size of Planck's constant, in which case the stationary phase approximation of the Feynman propagator can be justified. The validity of the diagonal approximation, on the other hand, can be granted if the classical dynamics is fully chaotic and if some sort of (energetic or disorder) averaging of the observables of interest is performed.

It is known, however, that there exist systematic and robust corrections beyond the diagonal approximation which may significantly affect quantum transport processes. These are, on the one hand, induced by coherent backscattering effects, the relevance of which for closed bosonic many-body systems has been studied in detail in Ref.~\cite{Engl2014PRL}. On the other hand, the apparent violation of norm conservation due to coherent backscatttering is compensated by loop contributions arising from Sieber-Richter trajectory pairs \cite{SR-pairs,quantumchaostransportKlaus,loops1,loops2,phd_mueller,loops3,loops4,loops5,one-leg-loops,one-leg-loops2}. The purpose of this section is to investigate these corrections in more detail for the bosonic many-body transport problem under consideration. We shall show in the end that they do not matter in the context of open atom-laser-like scattering systems.

\subsection{Semiclassical van Vleck--Gutzwiller theory for Bose--Hubbard systems}
Generally, the semiclassical approach can be understood as an expansion in a small parameter, namely the effective Planck constant $\hbar_{\mathrm{eff}}$ which vanishes in the classical limit. In a second quantized many-body theory, the role of this effective Planck constant is played by the inverse of the total number of particles $N$, \textit{i.e.} we identify $\hbar_{\mathrm{eff}}=1/N$. Applying the semiclassical theory to calculate the expectation value of some observable $\hat{O}$ then yields results of the form \cite{phd_mueller,phd_daniel}
\begin{equation}
  \langle\hat{O}\rangle(t)=O_{\mathrm{cl}}(t)+\sum_{k=1}^{\infty}\hbar_{\mathrm{eff}}^kO_{\mathrm{corr}}^{(k)}(t),
  \label{eq:semiclassical_expansion}
\end{equation}
where $O_{\mathrm{cl}}$ is the result of the corresponding classical theory and $O_{\mathrm{corr}}^{(k)}$ is the $k$th order quantum correction due to interference between certain classical paths, which can typically expressed in terms of $O_{\mathrm{cl}}$. In the case considered here, it will turn out that $O_{\mathrm{cl}}$ is given by the truncated Wigner result and all quantum corrections vanish.

The starting point of the semiclassical approach is the path integral representation of the propagator $\hat{K}(t)=\exp(-{i}\hat{H}t/\hbar)$ in a certain basis. For many-body systems, the most natural basis is constituted by the Fock states 
\begin{eqnarray}
\ket{\mathbf{n}} & \equiv & \ket{n_S} \otimes \ket{\dots,n_{-1},n_{0},n_{1},\dots} \nonumber \\
& = & \prod \limits_{\alpha} \frac{1}{\sqrt{n_{\alpha}!}} \left( \hat{\psi}_{\alpha}^\dagger\right)^{n_{\alpha}}\ket{0}
\end{eqnarray}
which are determined by the occupation numbers ({\it i.e.}~the numbers of 
atoms) $n_{\alpha}$ on the individual sites $\alpha$ (including the source) and 
which
are generated by applying the associated creation operators to the 
so-called vacuum state $\ket{0}$. 
However, it turns out that the semiclassical approximation for the propagator in Fock states also requires large occupations of each individual site \cite{Engl2014PRL,NoteFock} (typically $n_\alpha \ge 2$ \cite{Dujardin2015}) which is not granted in the scattering scenario considered here. 

A possibility to circumvent this problem is to use the \emph{quadrature} representation,
\begin{equation}
  K\left({\bf A}^{\mathrm{f}},{\bf A}^{\mathrm{i}};t\right)=\elemm{{\bf A}^{\mathrm{f}}}{e^{-i\hat{H}t/\hbar}}{{\bf A}^{\mathrm{i}}},
\label{eq:propagator_definition}
\end{equation}
where the quadrature states $\ket{\bf A}$ are defined by the eigenvalue equation \cite{VogelWelsch}
\begin{equation}
  \frac{1}{2}\left(\hat{\psi}_{\alpha}^{}+\hat{\psi}_\alpha^{\dagger}\right)\ket{{\bf A}}=A_{\alpha}\ket{{\bf A}}
\end{equation}
and can be expanded in Fock states according to
\begin{equation}
  \ket{{\bf A}}=\sum\limits_{\bf n}\left[\prod\limits_{\alpha}\frac{\exp\left(-A_{\alpha}^2\right)}{\sqrt{2^{n_{\alpha}}n_{\alpha}!\sqrt{\pi/2}}}H_{n_{\alpha}}\left(\sqrt{2}A_{n_{\alpha}}\right)\right]\ket{{\bf n}}.
\label{eq:quad2n}
\end{equation}
The analogy between these quadrature states and the position states of the harmonic oscillator is completed by the closure relation
\begin{equation}
  \hat{1}=\int\prod\limits_{\alpha}{\rm d}A_{\alpha}\ket{{\bf A}}\bra{{\bf A}}
  \label{eq:resolution_of_unity_quadratures}
\end{equation}
and their overlap
\begin{equation}
  \braket{{\bf A}}{{\bf A}^\prime}=\prod\limits_{\alpha}\delta\left(A_\alpha-A_\alpha^\prime\right).
\end{equation}
Finally, the action of an annihilation or creation operator on a quadrature state is given by
\begin{subequations}
  \begin{align}
    \hat{\psi}_{\alpha}\ket{{\bf A}}=&\left(A_{\alpha}+\frac{1}{2}\frac{\partial}{\partial A_{\alpha}}\right)\ket{{\bf A}} \\
    \hat{\psi}_{\alpha}^{\dagger}\ket{{\bf A}}=&\left(A_{\alpha}-\frac{1}{2}\frac{\partial}{\partial A_{\alpha}}\right)\ket{{\bf A}}.
  \end{align}
  \label{eq:creation_annihilation_quadrature}
\end{subequations}

If the total number of particles in the scattering region is large enough \cite{NoteLimits}, the path integral for the propagator in quadrature representation, given by Eq.~(\ref{eq:propagator_definition}), can be evaluated in a stationary phase approximation yielding the semiclassical propagator \cite{Gutzwiller,phd_tom}
\begin{equation}
  K \left({\bf A}^{\mathrm{f}},{\bf A}^{\mathrm{i}};t\right)\simeq \sum\limits_{\gamma}\mathcal{D}_\gamma\left({\bf A}^{\mathrm{f}},{\bf A}^{\mathrm{i}};t\right)e^{i R_\gamma\left({\bf A}^{\mathrm{f}},{\bf A}^{\mathrm{i}};t\right)/\hbar},
  \label{eq:semiclassical_propagator}
\end{equation}
where the sum runs over all trajectories $\gamma$ between $\ket{\mathbf{A}}$ and $\ket{\mathbf{A}'}$ that satisfy the classical equations of motion
\begin{equation}
  {i}\hbar\frac{\partial \psi_\alpha}{\partial t} (t)=\frac{\partial H_{\rm cl}\left({\boldsymbol\psi}^\ast,{\boldsymbol\psi}\right)}{\partial\psi_\alpha^\ast},
  \label{eq:eom}
\end{equation}
which are the same as in Eq.~(\ref{eq:StochAll}) with the classical Hamiltonian $H_{\rm cl}$ from Eq.~(\ref{eq:classicalhamiltonian}), and under the boundary conditions on the real parts of $\boldsymbol\psi$,
\begin{subequations}
  \begin{align}
    \Re\psi_\alpha(0)&=A_\alpha^{\mathrm{i}}, \\
    \Re\psi_\alpha(t)&=A_\alpha^{\mathrm{f}}.
  \end{align}
\end{subequations}
It is this sum together with the phase factor given by the action of the trajectory $\gamma$,
\begin{equation}
R_\gamma\left({\bf A}^{\mathrm{f}},{\bf A}^{\mathrm{i}};t\right)=\int_{0}^{t}{\rm d}s\left[2\hbar{\bf B}\cdot\frac{d}{ds}{\bf A}-H_{\rm cl}\left({\boldsymbol\psi}^\ast(s),{\boldsymbol\psi}(s)\right)\right],
\label{eq:action}
\end{equation}
that accounts for the interference effects lying beyond the truncated Wigner approach, the latter being, in the sense of $N\to\infty$, the classical limit of the quantum theory \cite{Dujardin2015AdP}. Here ${\bf A}(s)$ and ${\bf B}(s)$ are the real and imaginary part of ${\boldsymbol\psi}(s)$, respectively. Finally, the semiclassical amplitude is given by
\begin{eqnarray}
    \mathcal{D}_\gamma\left({\bf A}^{\mathrm{f}},{\bf A}^{\mathrm{i}};t\right)&=&\sqrt{\det\left(\frac{1}{-2\pi{i}\hbar}\frac{\partial^2R_\gamma\left({\bf A}^{\mathrm{f}},{\bf A}^{\mathrm{i}};t\right)}{\partial{\bf A}^{\mathrm{f}}\partial{\bf A}^{\mathrm{i}}}\right)} \nonumber \\
    &=&\sqrt{\det\left(\frac{1}{\pi{i}}\frac{\partial{\bf B}(0)}{\partial{\bf A}^{\mathrm{f}}}\right)},
  \label{eq:semiclassical_prefactor}
\end{eqnarray}
where the derivatives of the action with respect to the initial and final quadratures
\begin{subequations}
  \begin{align}
    \frac{\partial R_{\gamma}}{\partial {\bf A}^{\mathrm{i}}}&=-2\hbar{\bf B}(0),
    \label{eq:derivative_action_initial} \\
    \frac{\partial R_{\gamma}}{\partial {\bf A}^{\mathrm{f}}}&=2\hbar{\bf B}(t)
    \label{eq:derivative_action_final}
  \end{align}
  \label{eq:derivative_action}
\end{subequations}
have been used.

In order to proceed further, it is important to know the initial state \ket{{\boldsymbol\psi}^0} of the system. According to the considerations in section \ref{subsec:TW}, it will be assumed that the initial state is given by the vacuum in the waveguide and a coherent state in the source,
\begin{equation}
  \ket{{\boldsymbol\psi}^0}=\ket{\psi_S^0}\otimes\ket{0},
\end{equation}
where due to the gauge freedom $\psi_S^0=\sqrt{N}$ can be chosen to be real.

Inserting the resolution of unity (\ref{eq:resolution_of_unity_quadratures}), using the action of annihilation and creation operators on quadrature states (\ref{eq:creation_annihilation_quadrature}) as well as inserting the semiclassical approximation for the propagator (\ref{eq:semiclassical_propagator}) while letting act the derivatives of the propagator act on the exponential only \footnote{The derivative of the semiclassical prefactor yields terms of higher order of the effective Planck constant compared to those of the exponential, see \textit{e.g.}~\cite{Gutzwiller}.} then yields
\begin{eqnarray}
  \left\langle\hat{\psi}_{l^{}}^\dagger(t)\hat{\psi}_{l^\prime}^{}(t)\right\rangle&=&\left(\prod\limits_{\alpha}\sqrt{\frac{2}{\pi}}\int{\rm d}A_{\alpha}^{\mathrm{i}}\int{\rm d}A_{\alpha}^{\mathrm{i}^\prime}\int{\rm d}A_{\alpha}^{\mathrm{f}}\right) \nonumber \\
  &&\times\exp\Bigg\{-\sum\limits_{l^{\prime\prime}\in\mathbb{Z}}\left[\left(A_{l^{\prime\prime}}^{\mathrm{i}}\right)^2+\left(A_{l^{\prime\prime}}^{\mathrm{i}^\prime}\right)^2\right] \nonumber\\
  &&\qquad-\left(A_S^{\mathrm{i}}-\sqrt{N}\right)^2-\left(A_S^{\mathrm{i}^\prime}-\sqrt{N}\right)^2\Bigg\} \nonumber\\
  &&\times\sum\limits_{\substack{\gamma_1:{\bf A}^{\mathrm{i}}\to{\bf A}^{\mathrm{f}} \\ \gamma_2:{\bf A}^{\mathrm{i}^\prime}\to{\bf A}^{\mathrm{f}}}}\mathcal{D}_{\gamma_1}\mathcal{D}_{\gamma_2}^\ast\exp\left[\frac{i}{\hbar}\left(R_{\gamma_1}-R_{\gamma_2}\right)\right] \nonumber\\
  &&\qquad\times\left[\psi_{l^{}}^{(\gamma_2)\ast}(t)\psi_{l^\prime}^{(\gamma_1)}(t)-\frac{1}{2}\delta_{ll^\prime}\right],
  \label{eq:observable_semiclassical_start}
\end{eqnarray}
where Eq.~(\ref{eq:derivative_action_final}) has been used and the superscript $(\gamma_{1/2})$ indicates that the quantity is computed along the trajectory $\gamma_{1/2}$. The first exponential in Eq.~(\ref{eq:observable_semiclassical_start}) results from the overlap of quadrature states with the initial state and effectively reduces the integrations to a very small region, such that the initial quadratures ${\bf A}^{\mathrm{i}}, {\bf A}^{\mathrm{i}^\prime}$ lie very close to each other. Following standard semiclassical perturbation theory \cite{loschmidt_rodolfo}, we therefore replace ${\bf A}^{\mathrm{i}^\prime}$ by ${\bf A}^{\mathrm{i}}$ in the definition of the trajectories $\gamma_2$ and account for the induced error by expanding the action in the exponential up to linear order in the difference ${\bf A}^{\mathrm{i}^\prime}-{\bf A}^{\mathrm{i}}$.
The integration over ${\bf A}^{\mathrm{i}^\prime}$ is then simply a Gaussian integral, which can be performed exactly yielding
\begin{eqnarray}
    \lefteqn{\left\langle\hat{\psi}_{l^{}}^\dagger(t)\hat{\psi}_{l^\prime}^{}(t)\right\rangle\simeq
    \left(\prod\limits_{\alpha}2\int{\rm d}A_\alpha^{\mathrm{i}}\int{\rm d}A_\alpha^{\mathrm{f}}\right)\sum\limits_{\gamma_1,\gamma_2:{\bf A}\to{\bf A}^{\mathrm{f}}}\mathcal{D}_{\gamma_1}^{}\mathcal{D}_{\gamma_2}^\ast} \nonumber \\
    &&\times\exp\left\{-2\sum\limits_{\alpha\in\mathbb{Z}}\left[\left(A_\alpha^{\mathrm{i}}\right)^2+\frac{1}{4}\left(B_\alpha^{(\gamma_1)}(0)+B_\alpha^{(\gamma_2)}(0)\right)^2\right]\right\} \nonumber \\
    &&\times\exp\left\{-2\left|A_S^{\mathrm{i}}+\frac{i}{2}\left[B_S^{(\gamma_1)}(0)+B_S^{(\gamma_2)}(0)\right]-\sqrt{N}\right|^2\right\} \nonumber \\
    && \times\exp\left[\frac{i}{\hbar}\left(R_{\gamma_1}-R_{\gamma_2}\right)\right]\left[{\psi_{l^{}}^{(\gamma_2)}}^\ast(t)\psi_{l^\prime}^{(\gamma_1)}(t)-\frac{1}{2}\delta_{l^{}l^\prime}\right].
  \label{eq:observable_semiclassical}
\end{eqnarray}
This equation represents the starting point for the discussion of the diagonal approximation and the corresponding corrections provided by coherent backscattering and loop contributions.

\begin{figure}[t]
\centering
\subfigure[\label{fig:all_pairs}]{\includegraphics[width=0.45\columnwidth]{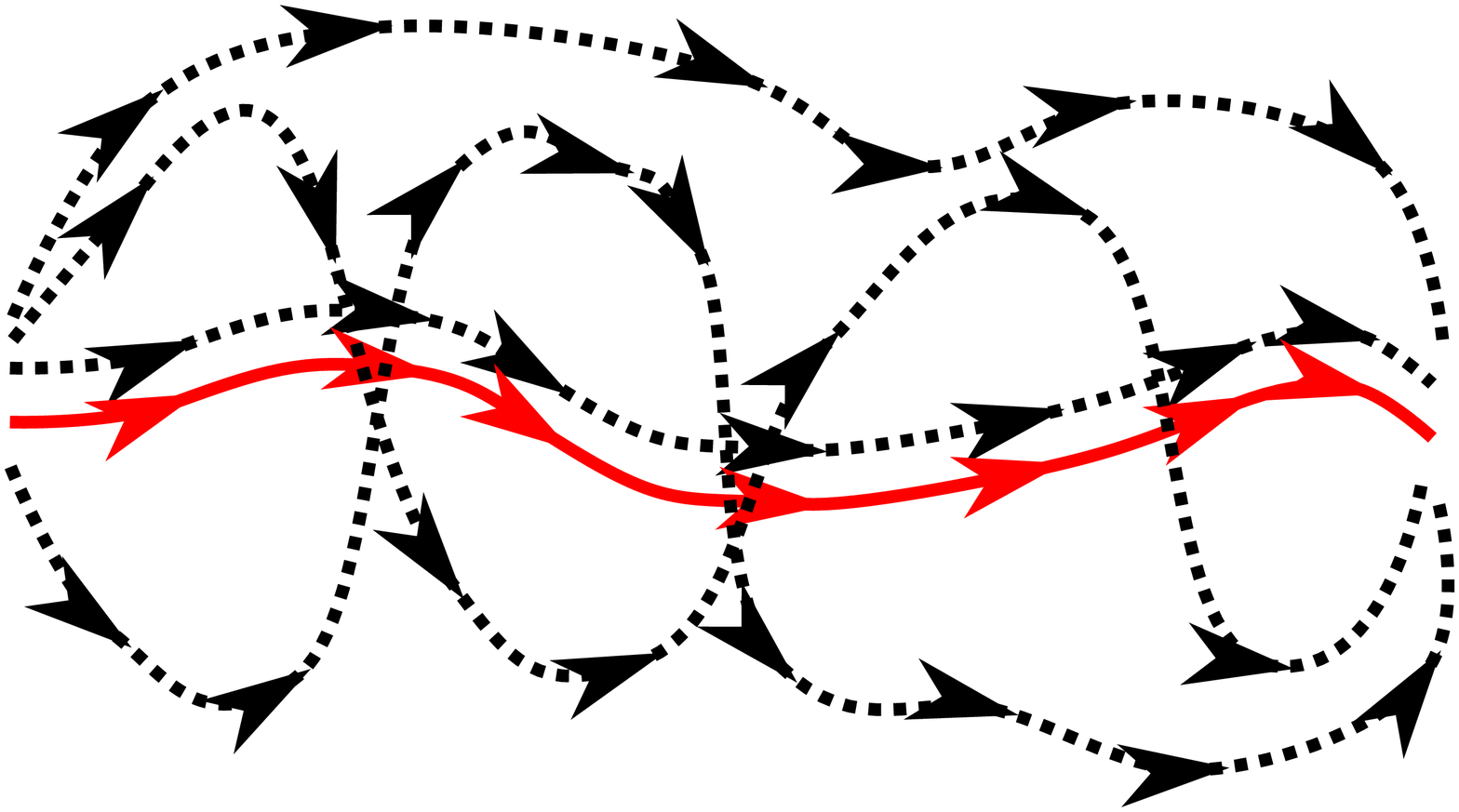}}
\hspace{0.03\textwidth}
\subfigure[\label{fig:diagonal}]{\raisebox{7mm}{\includegraphics[width=0.45\columnwidth]{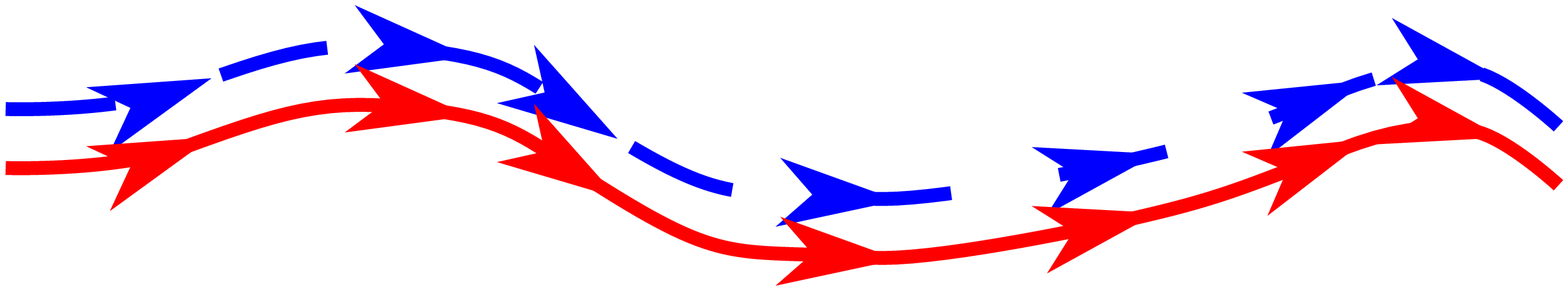}}} \\
\subfigure[\label{fig:cbs}]{\includegraphics[width=0.45\columnwidth]{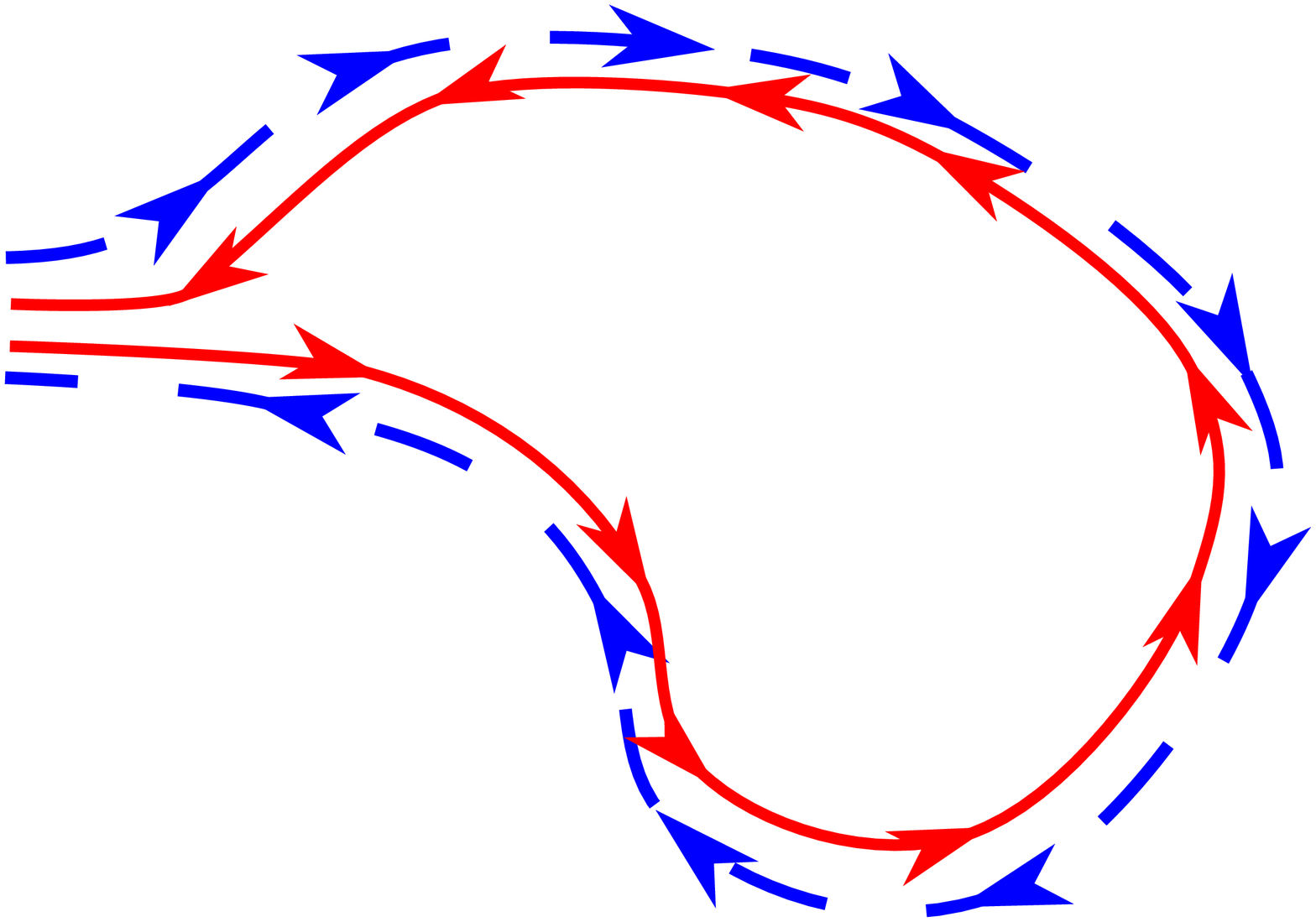}}
\hspace{0.03\textwidth}
\subfigure[\label{fig:loop}]{\includegraphics[width=0.45\columnwidth]{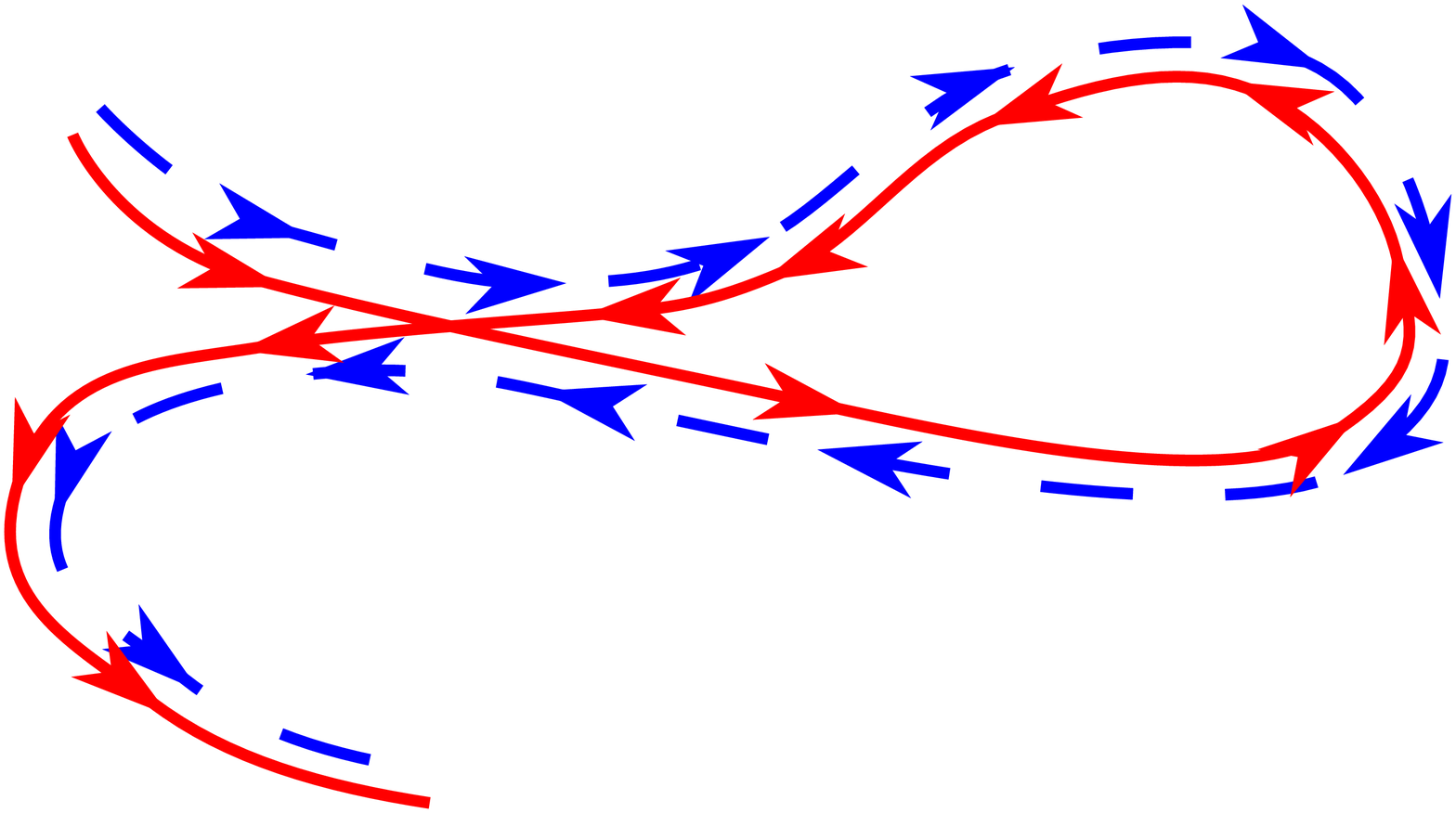}}
\caption{\label{fig:pairing}(color online) While for an individual system one would have to take into account all trajectories in the sum over partner orbits (\subref{fig:all_pairs}), under disorder average only certain partner trajectories are selected (\subref{fig:diagonal}-\subref{fig:loop}). These pairs are, ordered from largest to smallest contributions, the diagonal approximation (\subref{fig:diagonal}), coherent backscattering contributions (\subref{fig:cbs}) and loop contributions (\subref{fig:loop}).}
\end{figure}

\subsection{Closed systems}
Let us first consider the case of a closed disordered system with a finite number of sites, which could, \textit{e.g.}, describe an ultracold Bose gas that is contained within a finite optical lattice \cite{Engl2014PRL}.
When performing a disorder average, the action difference $R_{\gamma_1}-R_{\gamma_2}$ arising in Eq.~\eqref{eq:observable_semiclassical} varies strongly from one disorder realization to the next one, as long as the trajectories $\gamma_1$ and $\gamma_2$ are not correlated. Therefore, on average, most of the contributions to the double sum will cancel out and only pairs of trajectories that are close to each other during the whole evolution time will contribute. A short pictorial summary of the possible pairings is shown in Fig.~\ref{fig:pairing}.

For such pairs of trajectories, one can safely neglect the tiny differences in the prefactors $\mathcal{D}_{\gamma_j}$ and amplitudes $B_\alpha^{(\gamma_j)}(0)$ ($j=1,2$) between the two trajectories, such that only their action difference $\Delta R_{\gamma_1,\gamma_2}=R_{\gamma_1}-R_{\gamma_2}$ is taken into account. Equation~(\ref{eq:observable_semiclassical}) then becomes
\begin{eqnarray}
  \left\langle\hat{\psi}_{l^{}}^\dagger(t)\hat{\psi}_{l^\prime}^{}(t)\right\rangle
  &\simeq&
    \left(\prod\limits_{\alpha}2\int{\rm d}A_{\alpha}^{\mathrm{i}}\int{\rm d}A_{\alpha}^{\mathrm{f}}\right)\sum\limits_{\gamma:{\bf A}^{\mathrm{i}}\to{\bf A}^{\mathrm{f}}}\left|\mathcal{D}_{\gamma}^{}\right|^2 \nonumber \\
    &\times &\exp\left\{-2\sum\limits_{l\in\mathbb{Z}}\left[\left(A_l^{\mathrm{i}}\right)^2+\left(B_l^{(\gamma)}(0)\right)^2\right]\right\} \nonumber \\
    &\times &\exp\left\{-2\left(A_S^{\mathrm{i}}-\sqrt{N}\right)^2-2\left(B_S^{(\gamma)}(0)\right)^2\right\} \nonumber \\
    &\times &\left[{\psi_{l^{}}^{(\gamma)}}^\ast(t)\psi_{l^\prime}^{(\gamma)}(t)-\frac{1}{2}\delta_{l^{}l^\prime}\right]\nonumber \\
    &\times &\sum\limits_{\gamma_p}\exp\left(\frac{i}{\hbar}\Delta R_{\gamma,\gamma_p}\right).
  \label{eq:observable_semiclassical_paired}
\end{eqnarray}
Here the sum over $\gamma$ runs over all possible trajectories while the second sum over $\gamma_p$ includes all partner trajectories whose actions $R_{\gamma_p}$ are systematically correlated with the action $R_{\gamma}$ of the trajectory $\gamma$ and whose contributions thereby survive the disorder average.

\subsubsection{Diagonal approximation}
\label{subsec:da}
The easiest possibility for the partner trajectory $\gamma_p$ to yield a non-vanishing contribution on average is to choose it to be the same as $\gamma$: $\gamma_p=\gamma$. This is known as the diagonal approximation \cite{da} and has obviously a vanishing action difference. Then, by using the sum rule
\begin{equation}
  \left(\prod\limits_{\alpha}\int{\rm d}A_{\alpha}^{\mathrm{f}}\right)\sum\limits_{\gamma:{\bf A}\to{\bf A}^{\mathrm{f}}}\left|\mathcal{D}_\gamma\right|^2\ldots=\left(\prod\limits_{\alpha}\frac{1}{\pi}\int{\rm d}B_{\alpha}\right)\ldots,
  \label{eq:sum_rule}
\end{equation}
which is a direct consequence of $|\mathcal{D}_\gamma|^2=|\det(\partial{\bf B}^{(\gamma)}(0)/\partial{\bf A}^{\mathrm{f}})/\pi|$, the integration over ${\bf A}^{\mathrm{f}}$ can be transformed into an integration over ${\bf B}$.
This yields together with the integration over ${\bf A}^{\mathrm{i}}$ a Gaussian sampling over initial conditions, which are then evolved according to the equations of motion (\ref{eq:StochAll}). The observable is then calculated from the thereby evolved state only.
Thus with Eq.~(\ref{eq:sum_rule}) the shooting problem, where the real parts of ${\boldsymbol\psi}$ at initial and final time are fixed, is transformed into an initial value problem, which exactly yields the Truncated Wigner method described in Sec.~\ref{subsec:TW} \cite{Dujardin2015AdP}.

\subsubsection{Coherent Backscattering}
\label{subsec:cbs}
The second possibility for the partner trajectory $\gamma_p$ is to choose it to be the time reverse of $\gamma$, provided the latter is not identical to $\gamma$ (in which case the trajectory retraces itself in configuration space). Then obviously the action difference again vanishes. This kind of pairing requires that ${\bf A}^{\mathrm{f}}\simeq{\bf A}^{\mathrm{i}}$, {\it i.e.}~that the trajectory comes back to its initial point. As such trajectories can thus be paired with themselves and with their time-reversed counterparts, their contributions to Eq.~\eqref{eq:observable_semiclassical_paired} are therefore enhanced by a factor 2 as compared to other trajectories. This yields a coherent enhancement of backreflection \cite{Engl2014PRL}.

However, when considering single-particle observables such as the ones that we are focusing on here, the final quadrature ${\bf A}^{\mathrm{f}}$ is integrated over, which leads to a suppression of the coherent backscattering enhancement (or, more precisely, to a reduction of this enhancement by a factor that is of the order of the number of accessible many-body states with the same number of particles and about the same energy as the initial state \cite{Engl2014PRL}). This need not be the case for more sophisticated observables: the inverse participation ratio in the quantum many-body space, for instance, is enhanced by a factor 2 as compared to its classical value \cite{Engl2014PRL}. Mean on-site densities and inter-site currents, however, are not expected to be appreciable affected by coherent backscattering.

\subsubsection{Loop corrections}
Further possibilities for the partner trajectories have been found to be obtained by what is now known as loop contributions resulting from Sieber-Richter trajectory pairs \cite{phd_mueller,SR-pairs,quantumchaostransportKlaus,loops1,loops2,loops3,loops4,loops5,one-leg-loops,one-leg-loops2}, such as the one shown in Fig.~\ref{fig:loop}. These Sieber-Richter pairs consist of trajectories which in a certain very small region in phase space, the so-called encounter region, come close to their own time reverse. The partner trajectory closely follows then $\gamma$ or its time reverse all along the trajectory, except within the encounter region where it switches from following $\gamma$ to following its time reverse or vice versa. 

Such loop contributions are generally arising in quantum systems with a chaotic classical dynamics where the existence of partner trajectories with the above properties is granted by the shadowing theorem. While they turn out to significantly affect {\it e.g.}\ transport properties within mesoscopic systems \cite{quantumchaostransportKlaus,loops3,loops4,loops5,Daniel_nonlin_dynamics_nanosystems} and the Loschmidt echo \cite{loschmidt-echo_local_perturbation,fidelity_loops}, they can be shown to vanish for the evaluation of single-particle observables in ordinary time-evolution processes taking place within closed generic systems \cite{one-leg-loops,one-leg-loops2,continuity_equation_semiclassical}, in close analogy with the absence of coherent backscattering contributions. This also holds for the Bose--Hubbard systems that are considered here, which exhibit a second constant of motion in addition to the total energy, namely the total number of particles $N$. In Appendix \ref{appendix:loops} we show how to incorporate this additional conserved quantity.

\subsection{Open systems}
Open quantum many-body systems feature the complication that their number of classical degrees of freedom (\textit{i.e.}, the total number of Bose-Hubbard sites in our case) is infinitely large from a formal point of view, even if a discretization of the configuration space is employed.
While the diagonal approximation can still be applied in this context without any further considerations \cite{Dujardin2015AdP}, one has to be more careful with coherent backscattering and loop contributions.

On one hand, coherent backscattering contributions require the existence of trajectories that come back to their initial point, which is a very unlikely event to happen within an infinite system. Indeed, in our context of a guided atom laser, this would imply that \emph{all atoms} of the condensate come back to the reservoir after having explored the disordered region within the waveguide. Already from this point of view, one would expect that effects due to coherent backscattering in the many-body space are completely suppressed in open systems. On a more rigorous level, it can be shown (see Appendix \ref{appendix:splitting}) that in open systems the possibility of pairing a trajectory with its time-reversed counterpart requires self-retracing of the trajectory. This, however, implies that the trajectory would be identical with its time-reversed conterpart, which means that it does not contribute to coherent backscattering. Hence, coherent backscattering contributions do indeed vanish in the case of open systems.

A similar conclusion is obtained for the contributions of possible loop corrections within open systems. To investigate their relevance, it is convenient to employ a scattering approach in which the two semi-infinite left and right leads are formally separated from the finite scattering region which contains, as is illustrated in Fig.~\ref{fig:atlaser}(b), the disorder potential as well as the coupling to the source. As the interaction strength is vanishing outside the scattering region, the dynamics in the leads can be exactly integrated yielding integro-differential (\textit{i.e.}, temporally delayed) decay terms as well as quantum noise terms in the effective tW description of the scattering region \cite{Dujardin2014APB,Dujardin2015PRA}. We show in Appendix \ref{appendix:splitting} how to derive the corresponding semiclassical propagator describing the time evolution within the finite scattering system.

As the dynamics outside the scattering region is non-interacting,
the semiclassical approximation for the propagation in the leads is exact and can thus be represented by a single classical trajectory. 
Encounter regions where a trajectory comes close to its time-reversed counterpart in phase space may therefore arise only in the scattering region in which the presence of interactions can give rise to classical chaos  \cite{Lubasch2009,CBH1}. The resulting exponential sensitivity of the classical time evolution with respect to variations of the initial conditions is already sufficient for the existence of Sieber-Richter pairs \cite{SR-pairs}.
Their contributions to the time-dependent expectation values of single-particle observables can then again be evaluated following the standard theory of loops \cite{loops1,loops2,loops3,loops4,loops5,one-leg-loops,one-leg-loops2,continuity_equation_semiclassical,NoteErgodicity}. By virtue of the same arguments that are put forward in closed systems \cite{one-leg-loops,one-leg-loops2,continuity_equation_semiclassical}, we can again show that loop contributions vanish in the context of many-body scattering.

\section{Numerical results}
\label{sec:results}

\subsection{Anderson localization of weakly interacting atoms}
\label{subsec:alinteractingatoms}

We now apply the truncated Wigner method to investigate the transmission of the BEC across a disordered region of finite length $L$.
We specifically consider a Gaussian correlated disorder potential
that is generated in the same way as in Ref.~\cite{Paul2009PRA}, \textit{i.e.}
\begin{equation}
  \label{eq:GaussDis}
  V(x) = \frac{\hbar^2 A}{m} \int_0^L \frac{1}{\sqrt{2\pi}\sigma}\,e^{-\frac{(x-y)^2}{2\sigma^2}} \,\eta(y)\,dy.
\end{equation}
Here, the parameter $A$ controls the height of the disorder potential, $\sigma$ is the correlation length and $\eta(y)$ is a gaussian white noise with zero mean and unit variance, \textit{i.e.} $\langle \eta(x) \eta(y)\rangle = \delta(x-y)$ where $\langle\cdot\rangle$ denotes the average over different realizations of the disorder. The potential is constructed such that it vanishes on average, \textit{i.e.} $\langle V(x)\rangle=0$.
We shall choose the parameters $A=0.265 k^{3/2}$ and $k\sigma=\sqrt{2}$ in the following where we have $\mu = \hbar^2 k^2/(2m)$. The coupling to the source is tuned such that a steady-state density of $\rho^\varnothing=\sqrt{2}k$ would be obtained within the waveguide if there were no disorder nor interaction between the atoms.

\begin{figure}[t]
  \centering
  \includegraphics[width=\linewidth]{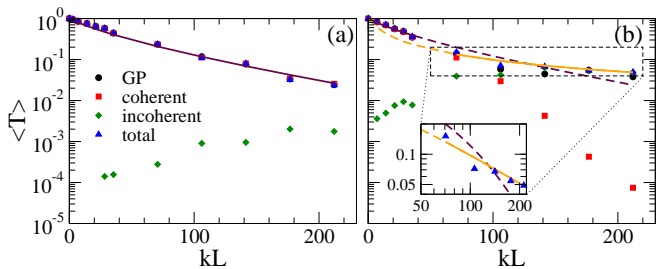}
  \caption{(color online) Transmission versus the length $L$ of the Gaussian-correlated disorder potential \eqref{eq:GaussDis} with the parameters $A=0.265 k^{3/2}$ and $k\sigma=\sqrt{2}$ for the atom-atom interaction strengths (a) $g=0.00095 \mu/k$ and (b) $g=0.048 \mu/k$.
The coherent, incoherent, and total transmissions obtained by the tW method are 
represented by (red) squares, (green) diamonds, and (blue) triangles, respectively, and the mean-field GP predictions for the transmission is shown by (black) circles. While for very weak interactions (a) the transmitted beam is fully coherent and subject to AL, a breakdown of AL is encountered at stronger interactions beyond a certain disorder length (b) yielding a fully incoherent beam in the transmitted region. The maroon solid lines show fits of Eq.~\eqref{eq:theodis} to the decay of the transmission within the coherent regime, yielding the localization lengths $k\Lloc\simeq 28$ in panel (a) and $k\Lloc\simeq 27$ in panel (b). The orange line in the latter panel shows a fit of a $1/(1+L/L_0)$ scaling of the transmission in the incoherent regime with $kL_0 \simeq 10.8$, which reproduces well the numerical data (as seen in the log-log plot in the inset).}
  \label{fig:Tlog}
\end{figure}

Fig.~\ref{fig:Tlog}(a) shows the coherent, incoherent, and total transmission of the condensate across the disorder potential \eqref{eq:GaussDis} as a function of its length $L$ in the presence of a rather weak interaction strength $g=0.00095 \mu/k$. We can directly notice that the coherent and total transmissions are nearly identical and agree very well with the transmission obtained from a mean-field GP calculation. The incoherent part of the transmission is suppressed by several orders of magnitude, but steadily increases with $L$, and we cannot exclude that it may eventually dominate the transport process for sufficiently long disorder regions.

A smoking gun of AL in disorder potentials is the dependence of the average transmission versus the length $L$ of the potential. More precisely, defining $\xi=L/\Lloc$, we have
\begin{equation}
  \label{eq:theodis}
  \langle  T(\xi) \rangle = \frac{e^{-\xi/4}}{2\sqrt{\pi \xi^3}}  \int_0^\infty du\, \frac{u^2}{\cosh(\frac{u}{2})} \exp\left(\frac{-u^2}{4\xi}\right),
\end{equation}
for the transmission of single particles across one-dimensional disordered regions \cite{Beenakker1994PRB},
where $\Lloc$ is defined as the \emph{localization length} which is proportional to the transport mean free path \cite{Beenakker1997RMP}. This formula was shown to be valid also in the mean-field limit described by the GP equation \cite{Paul2009PRA}, but does not take into account the possibility of depletion of the condensate beyond the mean-field approximation.

The maroon solid line in Fig.~\ref{fig:Tlog}(a) shows a least-mean-square fit of Eq.~\eqref{eq:theodis} to the total transmission, yielding very good agreement with the numerical data. The coherent flux of atoms therefore exhibits AL when passing through the gaussian-correlated disorder potential. Note that the numerically extracted value $k\Lloc\simeq 28$ of the localization length lies significantly below the theoretical value that would be predicted by the Born approximation \cite{Paul2009PRA,Lugan2009PRA}. This is not surprising since the latter requires $\hbar^2A/(m\mu k^{3/2})~\ll~1$ to be valid, which is not satisfied in our case.

\begin{figure}[t]
  \begin{center}
    \includegraphics[width=\linewidth]{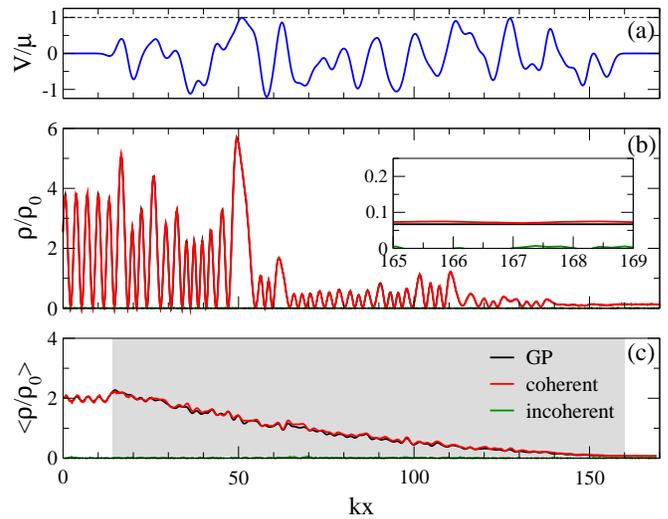}
    \caption{(color online) Density profiles for weak interaction $g=0.00095 \mu/k$ and a final propagation time of $\mu T/\hbar=1000$. Panel (a) shows one realization of the disorder potential generated according to Eq.\eqref{eq:GaussDis} and (b) represents the stationary density of an incoming beam of atoms going through the disorder potential represented in (a) with an inset magnifying the transmitted beam. The panel (c) shows the disorder average of the density profile taking into account 1000 realizations of the disorder. The gray background helps the reader to locate the disorder region. We can clearly see that the transport process is fully coherent and that the mean-field GP description (black line) agrees well with the truncated Wigner method (red line for the coherent and green line for the incoherent part).}
    \label{fig:dis_weak}
  \end{center}
\end{figure}

Fig.~\ref{fig:dis_weak} shows the corresponding
density profile of the atom laser beam for a fixed length of the disorder potential, namely $kL=141.42$, still in the case of very weak interaction. We can see that the GP and tW calculations yield nearly identical densities in the waveguide, both for individual disorder realizations [panel (b), the inset confirms that a stationary state has been reached] and on average [panel (c)]. One should point out that some fluctuations remain in the tW calculation, inherent to the stochastic nature of the technique.

\subsection{Breakdown of Anderson localization}
\label{subsec:breakdownofal}
We now focus our attention to the transition to incoherent transport. In Fig.~\ref{fig:Tlog}(b) we use the same parameter set as in Fig.~\ref{fig:Tlog}(a) except for the interaction strength which is increased to $g=0.048 \mu/k$. One can directly see that for small $kL$ we are still in the mean-field regime since the incoherent transmission is still several orders of magnitude below the coherent part. However, we now observe a rapid increase of the incoherent part of the transmission with $kL$. It becomes non-negligible at about $kL \simeq 100$, and for $kL>100$ the transport is fully incoherent. 

The dark (maroon) solid line in Fig.~\ref{fig:Tlog}(b) shows a fit of Eq.~\eqref{eq:theodis} to the numerically computed transmission data for $kL<50$ and the dark (maroon) dashed line extrapolates this fit in order to visualize what would happen if the transport remained fully coherent. 
The resulting localization length $k\Lloc\simeq 27$ appears to be slightly smaller than the one that was obtained in panel (a) for weaker atom-atom interaction. Indeed, the effect of a weak interaction can be accounted for in the theory of AL through replacing the wavenumber $k$ by an effective wavenumber $\tilde{k} = k\sqrt{1-1/(2\xi^2k^2)} < k$ taking into account the finite healing length $\xi=\hbar/\sqrt{2 m \rho g}$ of the propagating condensate \cite{Leboeuf2001PRA,Paul2007PRL,Paul2009PRA}, thus decreasing the localization length as seen in the results. 

While AL is therefore preserved for $kL<50$, we can see that for large disorder lengths $kL > 150$ the appearance of incoherent atoms allows for a higher transmission than predicted by the fit of Eq.~\eqref{eq:theodis}.
In this regime the decay of the transmission with $L$ is well fitted by a $1/(1+L/L_0)$ scaling with $k L_0 \simeq 10.8$ [light (orange) line in Fig.~\ref{fig:Tlog}(b)] which is characteristic for transport in systems with a loss of phase coherence between scattering events and which implies the breakdown of AL \cite{Paul2005PRA}. On the mean-field GP level, a non-stationary (turbulent) flow of atoms across the disordered region is expected in this regime \cite{Paul2005PRA,Ernst2010PRA}.
Interestingly, the GP and tW simulations are nevertheless in very good agreement, even though the coherence of the atom laser beam appears to be completely destroyed.

\begin{figure}[t]
  \begin{center}
    \includegraphics[width=\linewidth]{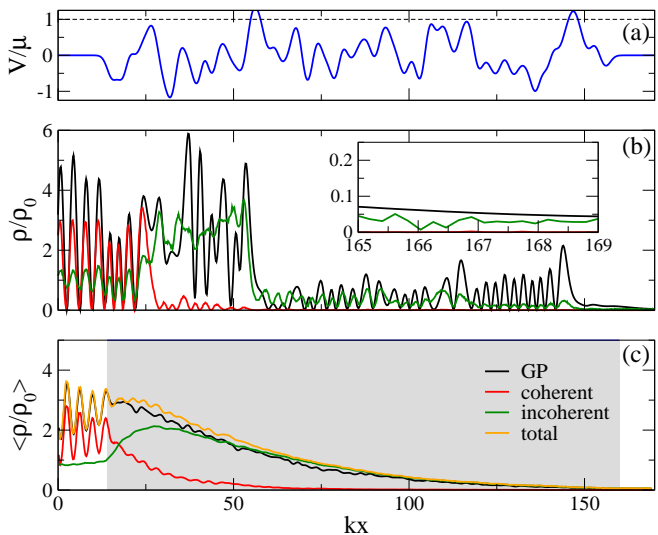}
    \caption{(color online) Density profiles in the non-localized regime for finite interaction $g=0.048 \mu/k$ and a final propagation time of $\mu T/\hbar=1000$. Panel (a) shows one realization of the disorder potential and panel (b) displays the corresponding density profile of an atom laser beam that propagates across this disorder potential. We can clearly see that the tW description of the transport process (red line for the coherent and green line for the incoherent part) features a breakdown of coherence and differs from the GP prediction (black line). The latter represents a ``snapshot'' of a non-stationary scattering process, which is reflected by the non-uniform GP density profile in the downstream region shown in the inset. Panel (c) displays the disorder average of the density profiles taking into account 1000 realizations. Despite the fact that the coherence is quickly lost within the disordered region (which is indicated by the shaded area), the disorder-averaged densities are nearly identical for both the GP and the tW method.}
    \label{fig:dis_finite}
  \end{center}
\end{figure}

This latter observation is confirmed in the comparison of the density profiles related to Fig.~\ref{fig:Tlog}(b) which are displayed Fig.~\ref{fig:dis_finite} for a disorder potential length of $kL=141.42$. Fig.~\ref{fig:dis_finite}(b) shows significantly different results predicted by the GP and tW calculations for an individual realization of the disorder. This illustrates that the mean-field description of the transport process is not valid any longer. Inspecting the inset one can furthermore infer that the GP simulation does not reach a stationary state, which is a consequence of the strong nonlinearity resulting from the atom-atom interaction, while the tW simulation attains a stationary density profile up to quantum fluctuations, featuring a non-negligible part of incoherent atoms. This shows that permanently time-dependent scattering processes in the mean-field limit amount to incoherent transport in tW calculations. However, as far as 
the disorder-averaged density profile is concerned, we see in Fig.~\ref{fig:dis_finite}(c) that the GP and tW results again coincide very well.

\section{Conclusions}
In summary, we studied in this work the one-dimensional transport of Bose--Einstein condensates across disorder potentials within a guided atom laser configuration. We used the truncated Wigner method for this purpose, which was recently adapted to deal with open systems \cite{Dujardin2015PRA}. Corrections beyond the mean-field Gross-Pitaveskii description of the transport process are mainly accounted for by means of quantum noise that results from the vacuum fluctuations within the waveguide \cite{Dujardin2015PRA}. From a semiclassical point of view, the tW method corresponds to the diagonal approximation within the framework of the van Vleck-Gutzwiller approach \cite{Dujardin2015AdP}. Systematic and robust corrections beyond this diagonal approximation can arise from coherent backscattering and loop contributions \cite{Engl2014PRL} but do not play any role in the transport context as we argued in this paper.

As for previous studies that were based on the mean-field GP description \cite{Paul2005PRA,Paul2007PRL,Paul2009PRA}, we find that Anderson localization is preserved for very weak interactions and/or short disordered regions. For stronger interaction a breakdown of AL is encountered, which is manifested by a crossover from an exponential to an algebraic $\propto 1/L$ decrease of the transmission with the disorder length $L$. This crossover is accompanied by a transition from a nearly perfectly coherent flux in the Anderson localized regime to a fully incoherent flux of atoms across the disordered region in the delocalized regime. We notice that this transition is remarkably well modeled by the mean-field GP method [see Figs.~\ref{fig:Tlog}(b) and \ref{fig:dis_finite}(b)] even though the latter is not valid in the incoherent regime. Evidently, this is a consequence of the disorder average that is performed. While an individual GP trajectory cannot correctly reproduce the many-body bosonic transport process due to the emergence of chaos in the non-stationary regime [see Fig.~\ref{fig:dis_finite}(a)], any averaging procedure removes the sensitivity of mean densities or transmissions on details of the chaotic dynamics and therefore yields rather similar results as a tW calculation that involves quantum noise. 

We finally point out that the comparison between the GP and the tW method may possibly lead to more subtle conclusions in transport problems of Bose-Einstein condensates that involve more than one spatial dimensions. Specifically for two-dimensional disordered systems it was shown that the presence of the nonlinearity in the GP equation may predict an inversion of the peak of coherent backscattering \cite{Hartung2008PRL} (which is here defined on the level of wave propagation in two spatial dimensions, and not in the Fock space of the many-body system) while a simple dephasing of the coherent backscattering peak in the presence of interaction is expected from a more microscopic point of view \cite{Geiger2012PRL,Geiger2013NJoP}. The tW method appears to be well suited to investigate this issue in more detail and to study other transport and localization phenomena of interacting bosonic matter waves in disordered systems, such as the Anderson transition in three spatial dimensions \cite{DelOrs14PRL}.

\begin{acknowledgments}
We thank Klaus Richter and Juan-Diego Urbina for fruitful discussions. Financial support from Fonds de la Recherche Scientifique de Belgique (F.R.S.-FNRS) and the DFG Forschergruppe FOR760 "Scattering Systems with Complex Dynamics" is gratefully acknowledged. This work was also financially supported  by a ULg research and mobility grant for T. E. The computational resources have been provided by the Consortium des \'{E}quipements de Calcul Intensif (C\'{E}CI), funded by the F.R.S.-FNRS under Grant No. 2.5020.11.
\end{acknowledgments}

\appendix
\section{Loop contributions in closed Bose-Hubbard systems}
\label{appendix:loops}

\begin{figure}[t]
\centering
\subfigure[\label{fig:two-leg-loop_1}]{\includegraphics[width=0.45\columnwidth]{2ll_1-2.eps}}
\hspace{0.03\textwidth}
\subfigure[\label{fig:one-leg-loop_1}]{\includegraphics[width=0.45\columnwidth]{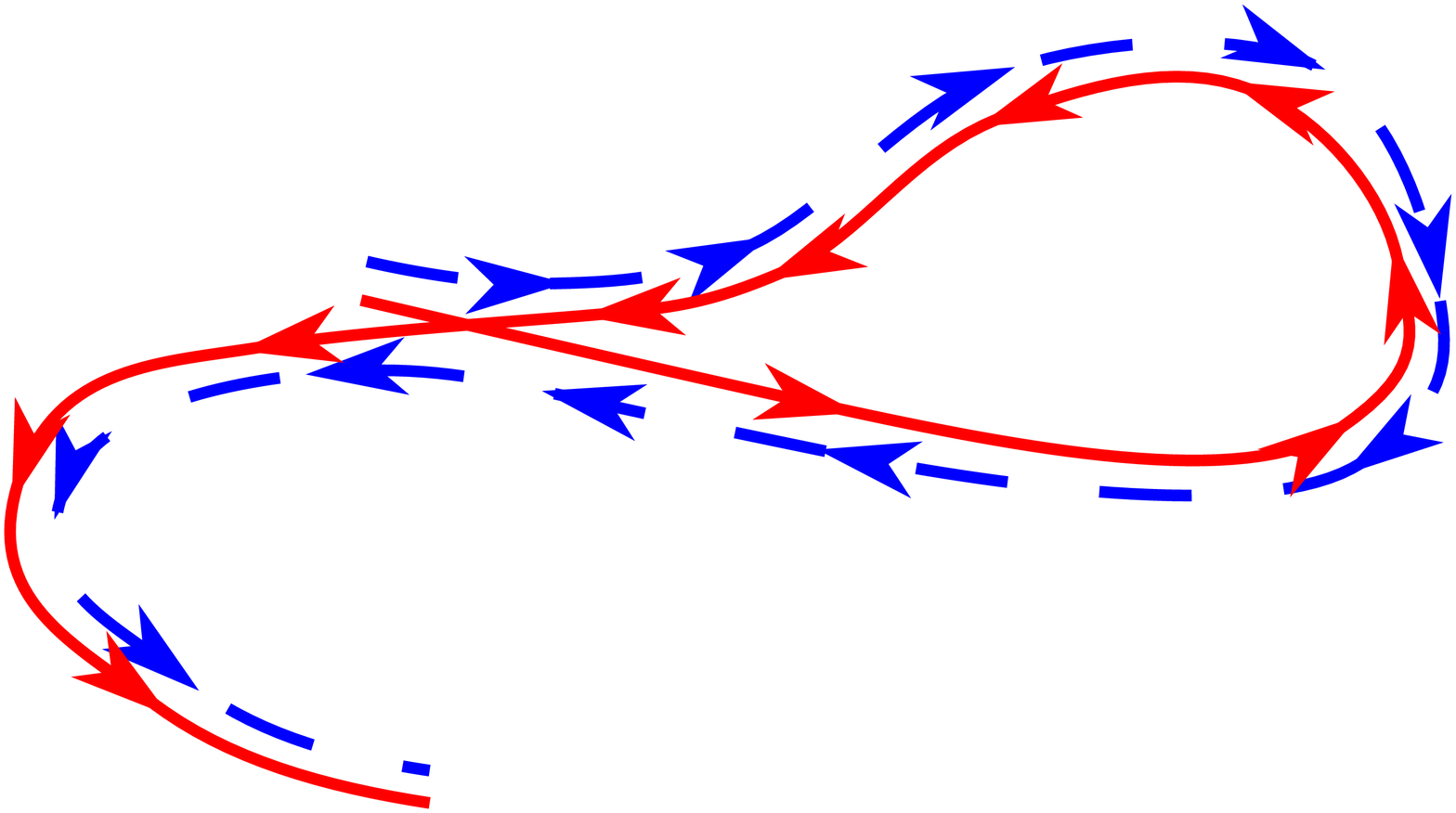}} \\
\subfigure[\label{fig:two-leg-loop_2}]{\includegraphics[width=0.45\columnwidth]{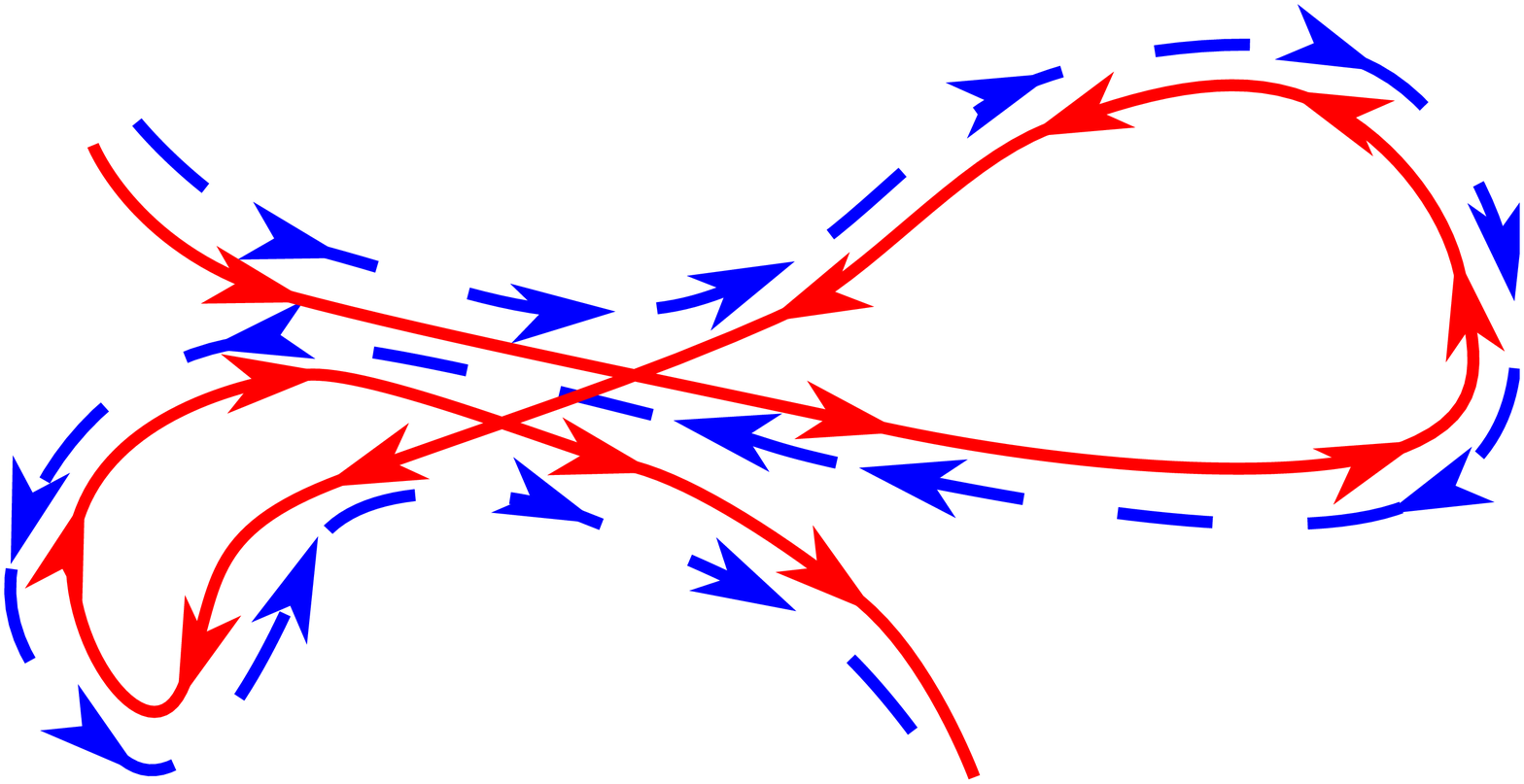}}
\hspace{0.03\textwidth}
\subfigure[\label{fig:one-leg-loop_2}]{\includegraphics[width=0.45\columnwidth]{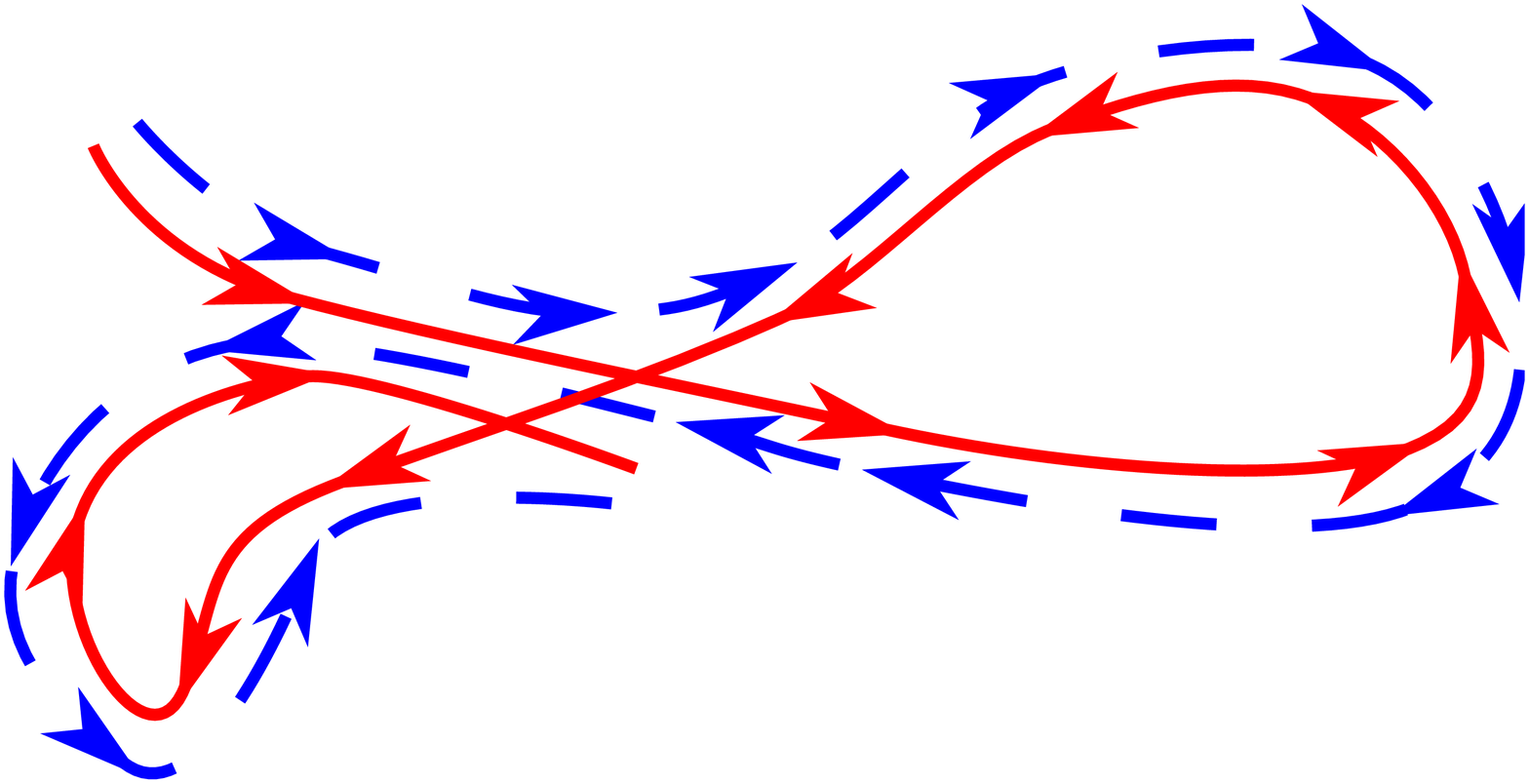}} \\
\subfigure[\label{fig:two-leg-loop_3}]{\includegraphics[width=0.3\columnwidth]{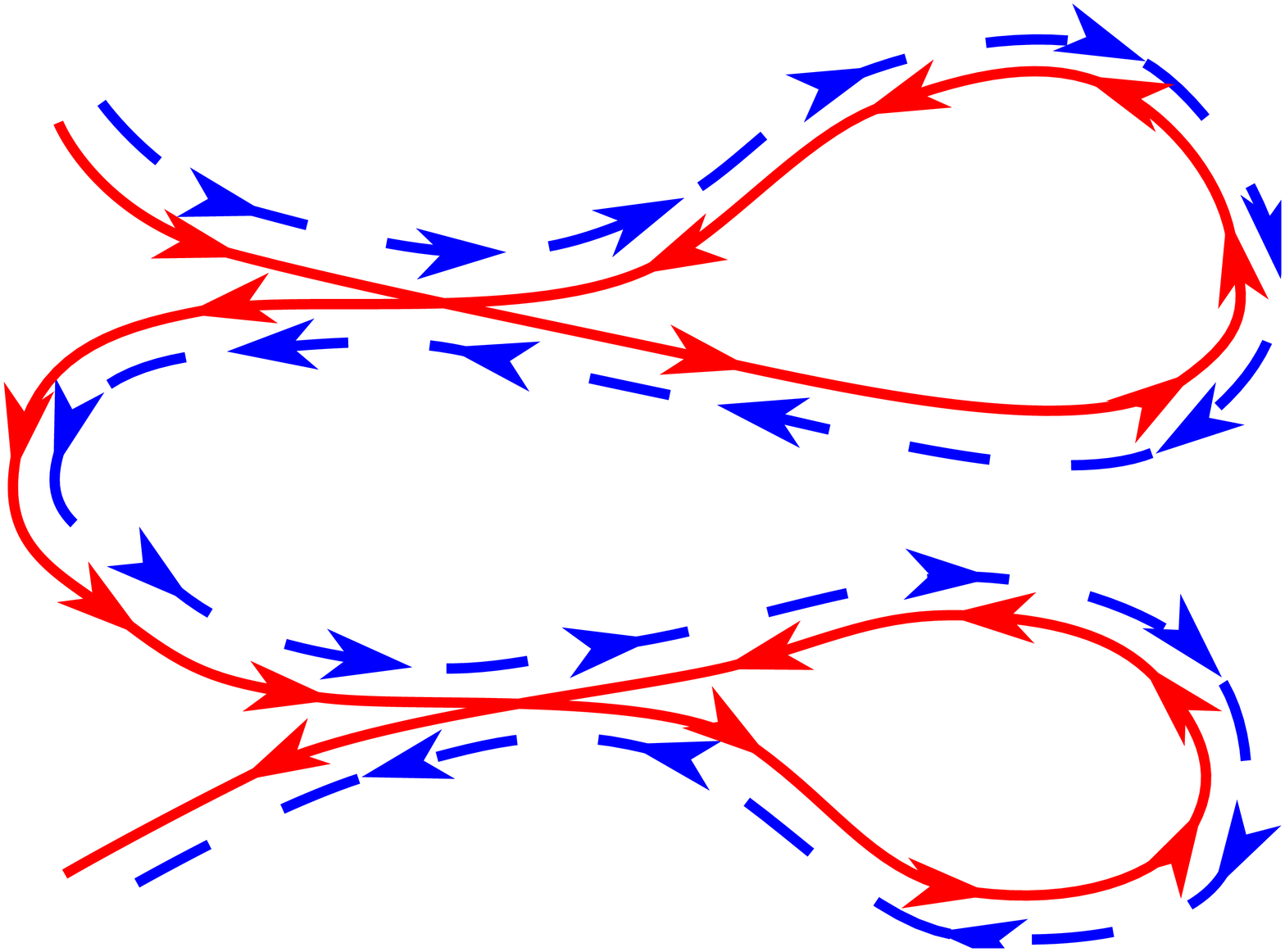}}
\hspace{0.01\textwidth}
\subfigure[\label{fig:one-leg-loop_3}]{\includegraphics[width=0.3\columnwidth]{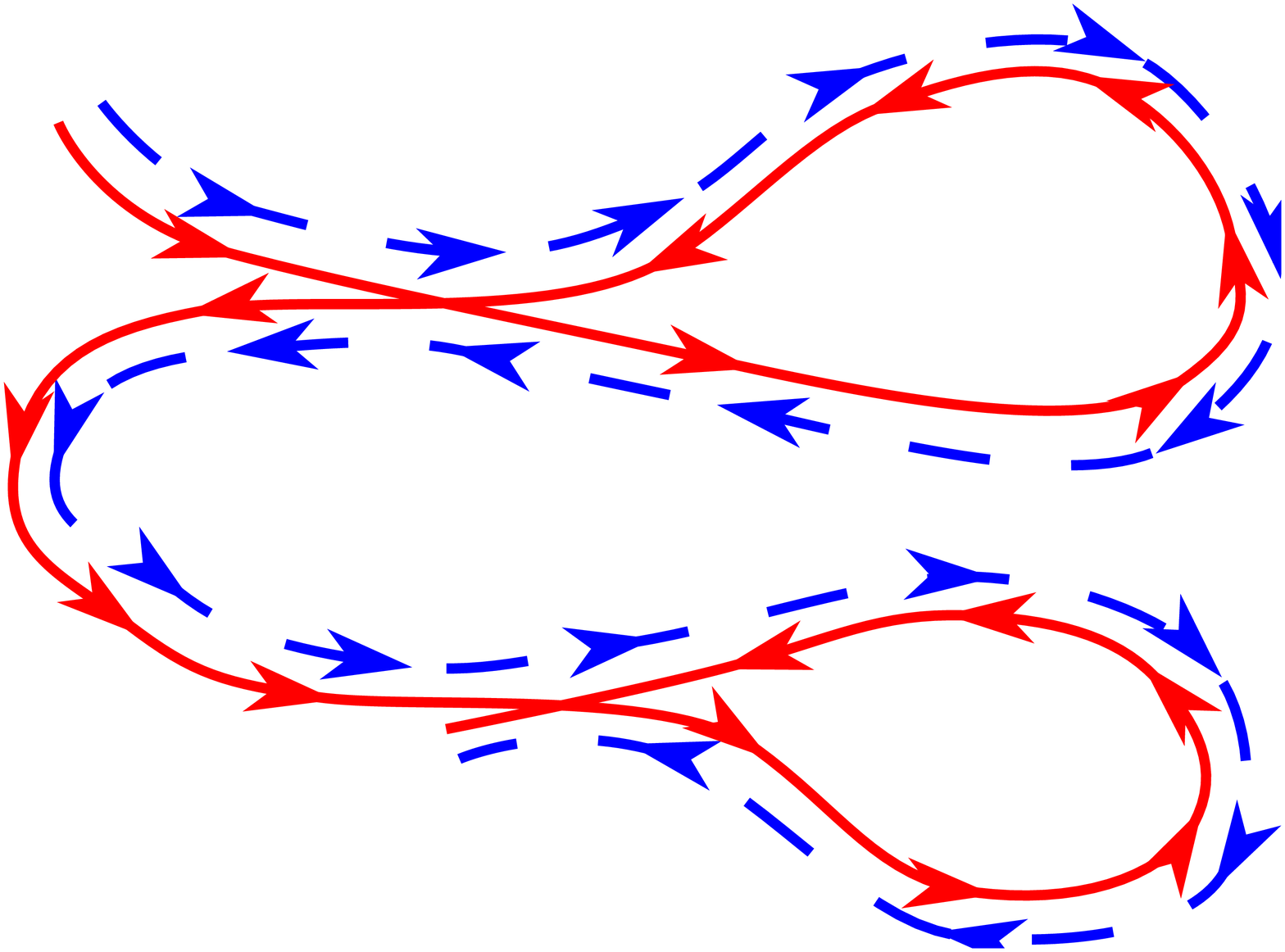}}
\hspace{0.01\textwidth}
\subfigure[\label{fig:no-leg-loop_3}]{\includegraphics[width=0.3\columnwidth]{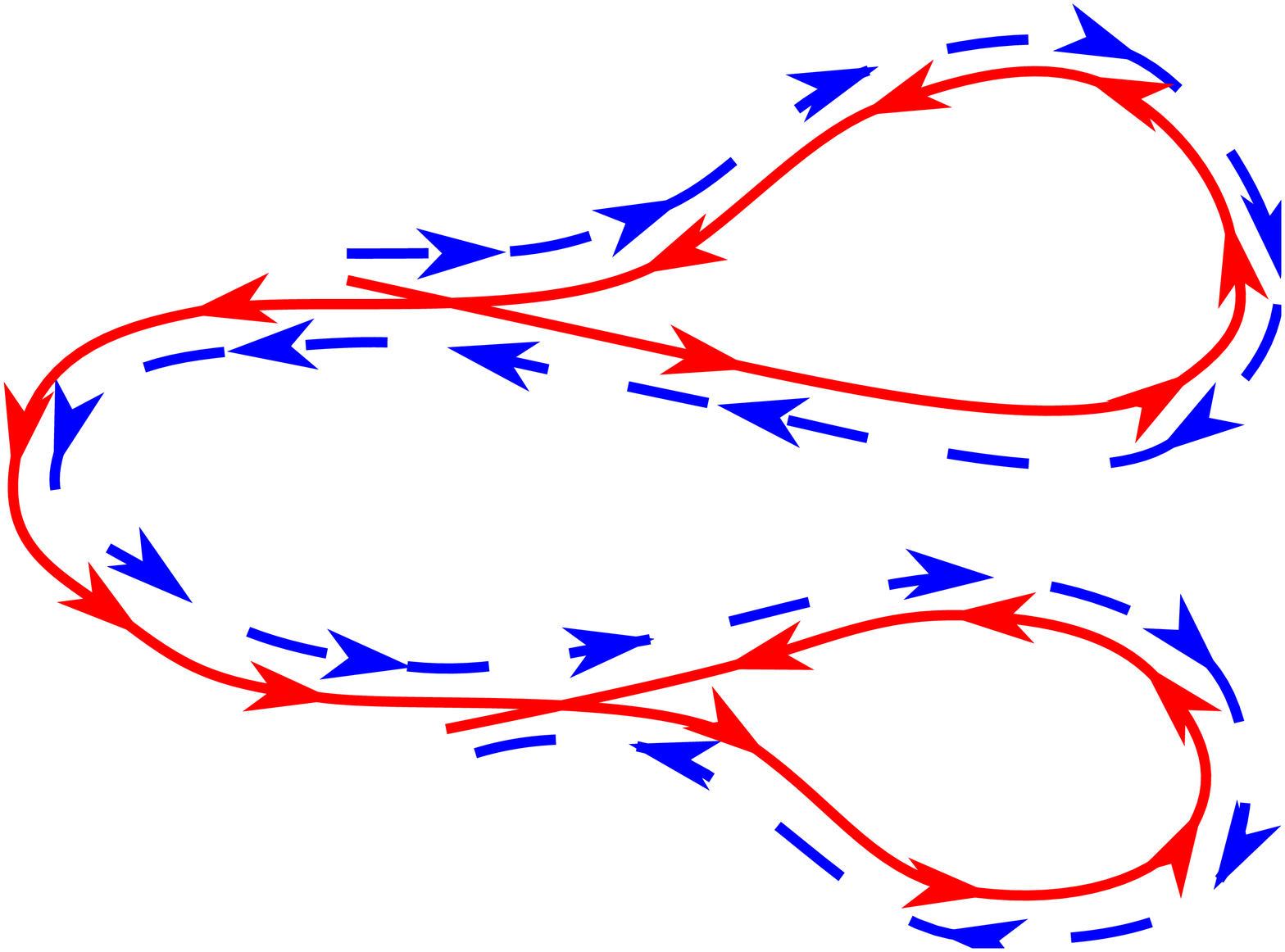}}
\caption{\label{fig:loops}(color online) Loop contributions up to second order. For the one-leg-loops (\subref{fig:one-leg-loop_1}, \subref{fig:one-leg-loop_2}, \subref{fig:one-leg-loop_3}) only one of the two possible cases of shifting one end into the encounter region is shown. Diagrams \subref{fig:two-leg-loop_1} and \subref{fig:one-leg-loop_1} are the leading order loop contributions, while the remaining ones are of next-to-leading order.}
\end{figure}

In this appendix we briefly explain how to evaluate loop contributions to Eq.~(\ref{eq:observable_semiclassical_paired}) for the specific case of a closed Bose-Hubbard system containing a finite number of $L$ sites and a finite number of $N$ particles.
Generally, such loop contributions originate from trajectories $\gamma$ that exhibit one or several nearby encounters in phase space with themselves or their time-reversed counterparts $\mathcal{T}\gamma$. Provided the classical dynamics is chaotic, there exist then partner trajectories $\gamma_p$ that follow pieces of $\gamma$ and/or $\mathcal{T}\gamma$ and, as is illustrated in Fig.~\ref{fig:loops}, switch from one piece to another one within the encounter region \cite{SR-pairs,quantumchaostransportKlaus,loops1,loops2,loops3,loops4,partner-trajectories,loops5,phd_mueller,one-leg-loops,one-leg-loops2,phd_daniel}. The overall contribution of the sum over the partner trajectories $\gamma_p$ in Eq.~(\ref{eq:observable_semiclassical_paired}) is then obtained by determining and counting all possible constructions of such partner trajectories and adding up their individual contributions \cite{partner-trajectories}.

We generally distinguish two-leg, one-leg, and no-leg loops, depending on how many (initial or final) ends of the trajectory lie outside the encounter regions. To evaluate their respective contributions, we follow
the standard literature on Sieber-Richter pairs \cite{loops1,loops2,loops3,loops4,partner-trajectories,loops5,phd_mueller,phd_daniel} where we replace the role of Planck's constant $\hbar$ by the effective one $\hbar_{\rm eff}=1/N$. We furthermore have to account for the presence of an additional constant of motion besides the total energy, namely the total number of particles $N$. This requires to perform the calculations in a reduced phase space with a fixed global phase and a fixed number of particles (see also \cite{phd_tom}).

The contribution of two-leg loops is then given by \cite{loops5,phd_daniel,phd_mueller}
\begin{eqnarray}
&&{\sum\limits_{\gamma_p}}^{(2ll)}\exp\left(\frac{i}{\hbar}\Delta R_{\gamma,\gamma_p}\right)= 
\sum\limits_{\bf v}\mathcal{N}\left({\bf v}\right)\int\limits_{-c}^{c}{\rm d}^{\left(L-2\right)\left(O-N_{\rm enc}\right)}s \nonumber \\
&& \qquad \times \int\limits_{-c}^{c}{\rm d}^{\left(L-2\right)\left(O-N_{\rm enc}\right)}u
\mathcal{P}\left({\bf s},{\bf u}, {\bf v}\right)\exp\left({i}N{\bf s}\cdot{\bf u}\right),
\label{eq:2ll-start}
\end{eqnarray}
where the sum runs over all partners giving rise to two-leg loops and ${\bf v}=\left(v_2,v_3,\ldots\right)$ is a vector whose entries are the numbers $v_o$ of $o$-encounters, with $o$ determined by counting how often the partner trajectory $\gamma_p$ differs from $\gamma$ within one self-encounter. Each term in the sum is weighted by the number $\mathcal{N}\left({\bf v}\right)$ of combinations of encounters for a given ${\bf v}$. Furthermore, the integrations run, roughly speaking, over the separations ${\bf s}$ and ${\bf u}$ along the stable and unstable manifolds, respectively, between the trajectories within each encounter, which determine the action difference $\Delta R_{\gamma,\gamma_p}=\hbar N{\bf s}\cdot{\bf u}$, with $c\ll N$ being some classical bound, which finally will drop out. The dimensionality of these integrations is determined by the total number $O=\sum_oov_o$ of traversals of encounters and the number $N_{\rm enc}=\sum_ov_o$ of encounters as well as the number $L-2$ of degrees of freedom obtained by fixing the energy $E$, which is determined by the energy of the trajectory $\gamma$, and the total number of particles $N$.

Finally
\begin{equation}
\mathcal{P}\left({\bf s},{\bf u},\bf{v}\right)=\frac{\left[t-\sum\limits_{\beta}o_\beta t_{\rm enc}^{(\beta)}\left({\bf s},{\bf u}\right)\right]^O}{O!\Omega_{N}^{O-N_{\rm enc}}\prod\limits_{\beta}t_{\rm enc}^{(\beta)}}
\label{eq:2ll-probability}
\end{equation}
is the probability that the trajectory $\gamma$ indeed exhibits the corresponding set of self-encounters, which follows from ergodicity arguments \cite{loops5,phd_daniel,phd_mueller}. Here,
\begin{equation}
t_{\rm enc}^{(\beta)}=\frac{1}{\lambda}\min\limits_{k^{},k^{\prime}\in\beta}\ln\frac{c^2}{\left|s_{k^{}}u_{k^{\prime}}\right|}
\end{equation}
is the encounter time of the $\beta$-th encounter with $\lambda$ being the Lyapunov exponent of the classical limit, where $s_k$ and $u_k$ denote the phase-space coordinates of the $k$th trajectory stretch along the stable and unstable manifold of the encounter, respectively (with $k\in\beta$ indicating that the $k$th trajectory stretch participates at the $\beta$-th encounter), and
\begin{eqnarray}
\Omega_N(E)&=&\int{\rm d}^{2L}\psi\delta\left(N-\sum\limits_{\alpha}\left(\left|\psi_{\alpha}\right|^2-\frac{1}{2}\right)\right) \nonumber\\
&& \times\delta\left(\theta_1\right)\delta\left(H_{\rm cl}\left({\boldsymbol\psi}^\ast,{\boldsymbol\psi}\right)-E\right)
\end{eqnarray}
is the phase space volume for fixed total number $N$ of particles, a fixed energy $E$ which is is determined by $\gamma$, and for an arbitrary choice of the global phase which is fixed by setting $\theta_1=0$.
Performing the integrations over the stable and unstable separations \cite{partner-trajectories} finally yields \cite{loops5,phd_daniel,phd_mueller,partner-trajectories}
\begin{eqnarray}
{\sum\limits_{\gamma_p}}^{(2ll)}\exp\left(\frac{i}{\hbar}\Delta R_{\gamma,\gamma_p}\right) & = &
\sum\limits_{\bf v}\mathcal{N}\left({\bf v}\right)
\left(\frac{2\pi}{N}\right)^{\left(L-2\right)\left(O-N_{\rm enc}\right)} \nonumber \\
&& \times \frac{t^{O-N_{\rm enc}}\prod\limits_{o}\left(-o\right)^{v_o}}{\left(O-N_{\rm enc}\right)!\Omega_N^{O-N_{\rm enc}}}.
\end{eqnarray}

If the trajectory starts within an encounter region, the corresponding encounter has to be treated differently.
Following Refs.~\cite{one-leg-loops,one-leg-loops2,phd_daniel}, the resulting one-leg loop contribution can be obtained by replacing in Eq.~(\ref{eq:2ll-start}) the sum over all possible combinations of encounters according to
\begin{equation}
\sum\limits_{\bf v}\mathcal{N}\left({\bf v}\right)\cdots=\sum\limits_{\bf v}\sum\limits_{o_1}v_{o_1}o_1\frac{\mathcal{N}\left({\bf v}\right)}{O}\cdots,
\end{equation}
to make the dependence on the first encounter explicit. Eq.~(\ref{eq:2ll-start}) then still holds provided we re-define
\begin{equation}
\mathcal{P}\left({\bf s},{\bf u},{\bf v}\right)=\int\limits_{0}^{\frac{1}{\lambda}\min\limits_{k\in1}\ln\frac{c}{\left|s_k\right|}}{\rm d}t^\prime\frac{\left[t-\sum\limits_{\beta}o_\beta t_{\rm enc}^{(\beta)}\left({\bf s},{\bf u}\right)\right]^{O-1}}{(O-1)!\Omega_{N}^{O-N_{\rm enc}}\prod\limits_{\beta}t_{\rm enc}^{(\beta)}},
\end{equation}
where the first encounter time is given by
\begin{equation}
t_{\rm enc}^{(1)}=t^\prime+\frac{1}{\lambda}\min\limits_{k\in1}\ln\frac{c}{\left|u_k\right|}.
\end{equation}
After again performing the integrations, one obtains \cite{one-leg-loops2,phd_daniel}
\begin{eqnarray}
&&{\sum\limits_{\gamma_p}}^{(1ll)}\exp\left(\frac{i}{\hbar}\Delta R_{\gamma,\gamma_p}\right)=
\sum\limits_{\bf v}\sum\limits_{o_1}o_1v_{o_1}\frac{\mathcal{N}\left({\bf v}\right)}{O} \nonumber \\ && \quad \times \left(\frac{2\pi}{N}\right)^{\left(L-2\right)\left(O-N_{\rm enc}\right)}
\frac{t^{O-N_{\rm enc}}\prod\limits_{\beta>1}\left(-o_\beta\right)}{\left(O-N_{\rm enc}\right)!\Omega_N^{O-N_{\rm enc}}}.
\end{eqnarray}

The same considerations hold if the trajectory ends instead of starts within an encounter region. Furthermore, for no-leg loops, where the trajectory both starts and ends within encounter regions [see Fig.~\ref{fig:loops}(g)], an analogous derivation yields \cite{one-leg-loops2,phd_daniel}
\begin{eqnarray}
&& {\sum\limits_{\gamma_p}}^{(0ll)}\exp\left(\frac{i}{\hbar}\Delta R_{\gamma,\gamma_p}\right)=
\sum\limits_{\bf v}\sum\limits_{o1,o_{N_{\rm enc}}}\mathcal{N}_{o_1,o_{N_{\rm enc}}}\left({\bf v}\right) \nonumber \\
&& \quad \times \left(\frac{2\pi}{N}\right)^{\left(L-2\right)\left(O-N_{\rm enc}\right)}
\frac{t^{O-N_{\rm enc}}\prod\limits_{\beta=2}^{N_{\rm enc}-1}\left(-o_\beta\right)}{\left(O-N_{\rm enc}\right)!\Omega_N^{O-N_{\rm enc}}},
\end{eqnarray}
with $N_{o_1,o_{N_{\rm enc}}}\left({\bf v}\right)$ being the number of combinations of encounters for given ${\bf v}$, if the first and last one are fixed to be an $o_1$- and $o_{N_{\rm enc}}$-encounter, respectively.

Summing up all these contributions is now a purely combinatorial problem and boils down to evaluating
$\mathcal{N}({\bf v}) (1 - 2 N_{\rm enc}/ O) + \sum_{o_1,o_{N_{\rm enc}}}\mathcal{N}_{o_1,o_{N_{\rm enc}}}/(o_1o_{N_{\rm enc}})$.
It has been shown in Ref.~\cite{continuity_equation_semiclassical} that this expression vanishes.
As a consequence, loop contributions do not play a role for the evaluation of Eq.~\eqref{eq:observable_semiclassical_paired} within closed systems.

\section{Effective propagator for the scattering region}
\label{appendix:splitting}
In this appendix we explain how to derive an effective van Vleck propagator within an open scattering region that is coupled to leads. To this end,
it is convenient to write the quantum Hamiltonian in the form
\begin{eqnarray}
\hat{H}&=&\hat{H}_{\rm scat}+\hat{H}_{\mathcal{L}}+\hat{H}_{\mathcal{R}} \nonumber \\
&-&\frac{E_{\delta}}{2}\left(\hat{\psi}_{0}^{\dagger}\hat{\psi}_{1}^{}+\hat{\psi}_{1}^{\dagger}\hat{\psi}_{0}^{}+\hat{\psi}_{L+1}^{\dagger}\hat{\psi}_{L}^{}+\hat{\psi}_{L}^{\dagger}\hat{\psi}_{L+1}^{}\right),
\end{eqnarray}
where in analogy with Refs.~\cite{Dujardin2014APB,Dujardin2015PRA} we assume that the scattering region (which is here denoted by the subscript `${\rm scat}$') consists of $L$ waveguide sites labeled by $1,\ldots,L$ and one source site labeled by $S$ (see Fig.~\ref{fig:atlaser}). The dynamics within the scattering region is governed by the Hamiltonian
\begin{eqnarray}
\hat{H}_{\rm scat}&=&\sum\limits_{l=1}^{L}\left[E_\delta+V_l-\mu+g_l\left(\hat{\psi}_l^{\dagger}\hat{\psi}_l^{}-1\right)\right]\hat{\psi}_l^{\dagger}\hat{\psi}_l^{} \\
&-&\sum\limits_{l=1}^{L-1}\frac{E_{\delta}}{2}\left(\hat{\psi}_l^{\dagger}\hat{\psi}_{l+1}+\hat{\psi}_{l+1}^{\dagger}\hat{\psi}_{l}\right)+\kappa\hat{\psi}_{l_S}^{\dagger}\hat{\psi}_S^{}+\kappa^{\ast}\hat{\psi}_S^{\dagger}\hat{\psi}_{l_S}^{}. \nonumber
\end{eqnarray}
while the Hamiltonians within the left and right leads are respectively given by
\begin{subequations}
\begin{align}
\hat{H}_{\mathcal{L}}=&\sum\limits_{l=-1}^{-\infty}\left[\left(E_\delta-\mu\right)\hat{\psi}_l^{\dagger}\hat{\psi}_l^{}-\frac{E_{\delta}}{2}\left(\hat{\psi}_l^{\dagger}\hat{\psi}_{l-1}+\hat{\psi}_{l-1}^{\dagger}\hat{\psi}_{l}\right)\right], \\
\hat{H}_{\mathcal{R}}=&\sum\limits_{l=L+1}^{\infty}\left[\left(E_\delta-\mu\right)\hat{\psi}_l^{\dagger}\hat{\psi}_l^{}-\frac{E_{\delta}}{2}\left(\hat{\psi}_l^{\dagger}\hat{\psi}_{l+1}+\hat{\psi}_{l+1}^{\dagger}\hat{\psi}_{l}\right)\right].
\end{align}
\end{subequations}

We now make the ansatz
\begin{equation}
\bra{{\bf A}}\exp\left(-\frac{{i}}{\hbar}\hat{H}t\right)\ket{{\bf A}^\prime}= \tilde{K}\left({\bf A},{\bf A}^\prime,t\right)\prod\limits_{j\in\{\mathcal{L},\mathcal{R}\}}K_{j}\left({\bf A},{\bf A}^{\prime},t\right),
\end{equation}
with the free propagators for the leads
\begin{equation}
K_{\mathcal{L}/\mathcal{R}}\left({\bf A},{\bf A}^{\prime},t\right)=\bra{{\bf A}_{\mathcal{L}/\mathcal{R}}}\exp\left(-\frac{{i}}{\hbar}\hat{H}_{\mathcal{L}/\mathcal{R}}t\right)\ket{{\bf A}'_{\mathcal{L}/\mathcal{R}}}
\end{equation}
where $\ket{{\bf A}_{\mathcal{L}/\mathcal{R}}}$ denote the quadrature states for the leads. A path integral representation for $\tilde{K}$ can then be found according to
\begin{widetext}
\begin{eqnarray}
\tilde{K}\left({\bf A},{\bf A}^{\prime},t\right)&=&\lim\limits_{M\to\infty}\int{\rm d}^{L+1}\tilde{A}^{(1)}\int\frac{{\rm d}^{L+1}\tilde{B}^{(1)}}{\left(4\pi\right)^{L+1}}\cdots\int{\rm d}^{L+1}\tilde{A}^{(M-1)}\int\frac{{\rm d}^{L+1}\tilde{B}^{(M-1)}}{\left(4\pi\right)^{L+1}}\int\frac{{\rm d}^{L+1}\tilde{B}^{(M)}}{\left(4\pi\right)^{L+1}} \nonumber \\
&&\times\exp\left\{\frac{{i} t}{4\hbar M}\sum\limits_{m=1}^{M}\left[2\hbar\tilde{{\bf B}}^{(m)}\cdot\frac{\tilde{{\bf A}}^{(m)}-\tilde{{\bf A}}^{(m-1)}}{t/M}-\tilde{H}_{\rm cl}\left(\tilde{\boldsymbol\psi}^{(m)\ast},\tilde{\boldsymbol\psi}^{(m)},mt/M\right)\right]\right\}
\label{eq:pathintegral_splitted}
\end{eqnarray}
with $\tilde{\boldsymbol\psi}^{(m)}=\tilde{\bf A}^{(m-1)}+{i}\tilde{\bf B}^{(m)}$, $\tilde{\bf A}^{(0)}=\tilde{\bf A}^{\prime}$ and $\tilde{\bf A}^{(M)}=\tilde{\bf A}$, where the tilde indicates that the concerned amplitudes and quadratures are restricted to the scattering region. The classical Hamiltonian of the scattering region is given by
\begin{eqnarray}
\tilde{H}_{\rm cl}\left(\tilde{\boldsymbol\psi}^{(m)\ast},\tilde{\boldsymbol\psi}^{(m)},mt/M\right)&=& 
H_{\rm cl}\left(\tilde{\boldsymbol\psi}^{(m)\ast},\tilde{\boldsymbol\psi}^{(m)}\right) \nonumber \\
&+&E_\delta{\bf A}_{\mathcal{L}}\Im\left[U_{\mathcal{L}}^{\rm T}(t)\right]\Im\left[U_{\mathcal{L}}^{\dagger}(mt/M){\bf e}^{(\mathcal{L})}\tilde{\psi}_0^{(m)}\right]-E_\delta{\bf A}_{\mathcal{L}}^\prime\Im\left[U_{\mathcal{L}}(t)\right]\Im\left[U_{\mathcal{L}}^{\dagger}(t-mt/M){\bf e}^{(\mathcal{L})}\tilde{\psi}_0^{(m)}\right] \nonumber \\
&+&E_\delta{\bf A}_{\mathcal{R}}\Im\left[U_{\mathcal{R}}^{\rm T}(t)\right]\Im\left[U_{\mathcal{R}}^{\dagger}(mt/M){\bf e}^{(\mathcal{R})}\tilde{\psi}_L^{(m)}\right]-E_\delta{\bf A}_{\mathcal{R}}^\prime\Im\left[U_{\mathcal{R}}(t)\right]\Im\left[U_{\mathcal{R}}^{\dagger}(t-mt/M){\bf e}^{(\mathcal{R})}\tilde{\psi}_L^{(m)}\right] \nonumber \\
&+&\frac{tE_{\delta}^2}{2M\hbar}\sum\limits_{m^\prime=1}^{m-1}\Im\left[\tilde{\psi}_{0}^{(m)\ast}{\bf e}^{(\mathcal{L})\rm T}U_{\mathcal{L}}^{\dagger}(t-mt/M)\right]\Im\left[U_{\mathcal{L}}^{\rm T}(t)\right]\Im\left[U_{\mathcal{L}}^{\dagger}(m^\prime t/M){\bf e}^{(\mathcal{L})}\tilde{\psi}_{0}^{(m^\prime)}\right] \nonumber \\
&+&\frac{tE_{\delta}^2}{2M\hbar}\sum\limits_{m^\prime=1}^{m-1}\Im\left[\tilde{\psi}_{L}^{(m)\ast}{\bf e}^{(\mathcal{R})\rm T}U_{\mathcal{R}}^{\dagger}(t-mt/M)\right]\Im\left[U_{\mathcal{R}}^{\rm T}(t)\right]\Im\left[U_{\mathcal{R}}^{\dagger}(m^\prime t/M){\bf e}^{(\mathcal{R})}\tilde{\psi}_{L}^{(m^\prime)}\right],
\end{eqnarray}
\end{widetext}
which parametrically depends on the initial and final quadratures ${\bf A}_{\mathcal{L}}$ and ${\bf A}_{\mathcal{R}}$ within the leads. Here,
\begin{eqnarray}
  H_{\textrm{cl}}(\boldsymbol{\psi},\boldsymbol{\psi}^*) &=& \sum_{l=1}^{L}\left[ (E_\delta + V_l-\mu-g_l)|\psi_l|^2 + \frac{g_l}{2} |\psi_l|^4\right]\nonumber \\
  && - \frac{E_\delta}{2}\sum_{l=1}^{L-1}\left[\left(\psi^*_{l+1}\psi_l + \psi^*_l\psi_{l+1}\right) \right] \nonumber \\
  && + \kappa^*(t)\psi_S^*\psi_{l_S} + \kappa(t)\psi^*_{l_S}\psi_S ,
\end{eqnarray}
is the classical Hamiltonian in the scattering region without the leads, and we define
\begin{subequations}
\begin{align}
{\bf e}^{(\mathcal{L})}&=\left( e^{(\mathcal{L})} \right)_{l\leq0},
& e_{0}^{(\mathcal{L})}=1, && e_{l\neq0}^{(\mathcal{L})}=0, \\
{\bf e}^{(\mathcal{R})}&=\left( e_{l}^{(\mathcal{R})} \right)_{l>L},
& e_{L+1}^{(\mathcal{R})}=1, && e_{l\neq L+1}^{(\mathcal{R})}=0.
\end{align}
\end{subequations}
$U_{\mathcal{L}/\mathcal{R}}(t)$ are the classical (single-particle) propagators for the leads, {\it i.e.}~the (unitary) solutions to the equations
\begin{subequations}
\begin{align}
{i}\hbar\frac{\partial U_{\mathcal{L}}(t)}{\partial t}&=H_{\rm cl}^{(\mathcal{L})}U_{\mathcal{L}}(t), & U_{\mathcal{L}}(0)&=1, \\
{i}\hbar\frac{\partial U_{\mathcal{R}}(t)}{\partial t}&=H_{\rm cl}^{(\mathcal{R})}U_{\mathcal{R}}(t), & U_{\mathcal{R}}(0)&=1,
\end{align}
\end{subequations}
where $H_{\rm cl}^{(\mathcal{L}/\mathcal{R})}$ denote the single-particle Hamiltonian matrices within the left and right lead, respectively. They have the matrix elements
\begin{align}
\left(H_{\rm cl}^{(\mathcal{L}/\mathcal{R})}\right)_{l,l^\prime}&=
\left(E_{\delta}-\mu\right)\delta_{l^{},l^{\prime}}-\frac{E_{\delta}}{2}\left(\delta_{l^{},l^{\prime}+1}+\delta_{l^{},l^{\prime}-1}\right),
\end{align}
with $l^{},l^{\prime}\in\{\ldots,-2,-1,0\}$ for the left and $l^{},l^{\prime}\in\{L+1,L+2,L+3,\ldots\}$ for the right lead.

The semiclassical approximation is then obtained by evaluating the integrals in Eq.~(\ref{eq:pathintegral_splitted}) using the stationary phase approximation and finally taking the limit $M\to\infty$ of an infinite number of intermediate time steps. This yields the effective van Vleck propagator
\begin{equation}
\tilde{K}\left({\bf A},{\bf A}^{\prime};t\right)=\sum\limits_{\gamma:\tilde{{\bf A}}^{}\to\tilde{{\bf A}}^{\prime}}\tilde{\mathcal{D}}_\gamma\left({\bf A},{\bf A}^{\prime},t\right)\exp\left[\frac{i}{\hbar}\tilde{R}_\gamma\left({\bf A},{\bf A}^{\prime},t\right)\right],
\end{equation}
where the trajectories $\gamma$ are determined by the equations of motion
\begin{equation}
{i}\hbar\dot{\tilde{\psi}}_l(t)=\frac{\partial \tilde{H}_{\rm cl}\left(\tilde{\boldsymbol\psi}^\ast(t),\tilde{\boldsymbol\psi}(t),t\right)}{\partial\tilde{\psi}_l^\ast(t)},
\end{equation}
[which are now integro-differential equations that, however, can still be transformed into Eq.~(\ref{eq:StochAll})] and by the boundary conditions
\begin{subequations}
\begin{align}
\Re\tilde{\psi}_l(0)&=\tilde{A}_l^{\prime}, \\
\Re\tilde{\psi}_l(t)&=\tilde{A}_l^{}
\end{align}
\end{subequations}
on the real parts of $\tilde{\psi}_l$ with $l\in\{S,1,\ldots,L\}$.
The action of the trajectory is given by
\begin{equation}
\tilde{R}_\gamma\left({\bf A}^{\mathrm{f}},{\bf A}^{\mathrm{i}},t\right)=
\int\limits_{0}^{t}{\rm d}t^\prime\left[2\hbar\tilde{{\bf B}}\cdot\dot{\tilde{{\bf A}}}-\tilde{H}_{\rm cl}\left(\tilde{\boldsymbol\psi}^\ast(t^\prime),\tilde{\boldsymbol\psi}(t^\prime),t^\prime\right)\right],
\end{equation}
where, again, $\tilde{{\bf A}}(t^\prime)$ and $\tilde{{\bf B}}(t^\prime)$ denote the real and imaginary parts of $\tilde{\boldsymbol\psi}(t^\prime)$, respectively. Finally, the semiclassical amplitude is given by
\begin{equation}
\tilde{\mathcal{D}}_\gamma\left({\bf A}^{},{\bf A}^{\prime},t\right)=\sqrt{\det\left(\frac{1}{-2\pi{i}\hbar}\frac{\partial^2\tilde{R}_\gamma\left({\bf A}^{},{\bf A}^{\prime},t\right)}{\partial\tilde{{\bf A}}^{}\partial\tilde{{\bf A}}^{\prime}}\right)},
\end{equation}
where the tilde indicates that the derivatives are taken with respect to the components corresponding to the scattering region only.
Since the exponential in the path integral for the propagators within the leads is quadratic in the fields, its corresponding semiclassical approximation is exact and yields
\begin{widetext}
\begin{equation}
K_{\mathcal{L}/\mathcal{R}}\left({\bf A},{\bf A}^{\prime},t\right)=\frac{\exp\left[-\frac{i}{4}\left(\begin{array}{c}{\bf A}_{\mathcal{L}/\mathcal{R}}^{\prime} \\
{\bf A}_{\mathcal{L}/\mathcal{R}}\end{array}\right)\left(\begin{array}{cc}\Im\left[U_{\mathcal{L}/\mathcal{R}}(t)\right]^{-1}\Re\left[U_{\mathcal{L}/\mathcal{R}}(t)\right] & -\Im\left[U_{\mathcal{L}/\mathcal{R}}(t)\right]^{-1} \\ -\Im\left[U_{\mathcal{L}/\mathcal{R}}^{\rm T}(t)\right]^{-1} & \Re\left[U_{\mathcal{L}/\mathcal{R}}(t)\right]\Im\left[U_{\mathcal{L}/\mathcal{R}}(t)\right]^{-1}\end{array}\right)\left(\begin{array}{c}{\bf A}_{\mathcal{L}/\mathcal{R}}^{\prime} \\ {\bf A}_{\mathcal{L}/\mathcal{R}}\end{array}\right)\right]}{\sqrt{\det\left[-4\pi{i}U_{\mathcal{L}/\mathcal{R}}(t)\Im\left[U_{\mathcal{L}/\mathcal{R}}(t)\right]^{\rm T}\right]}}.
\end{equation}
\end{widetext}
By construction, the pairing of trajectories takes then place in the scattering region only.

The fact that the total number of particles is not conserved allows us to deal with the coherent backscattering contribution on the level of the quadrature representation. However, when pairing a trajectory with its time reversed counterpart, one needs ${\bf A}^{\prime}\simeq{\bf A}^{}$. Hence, we replace again ${\bf A}^{\prime}$ by ${\bf A}^{}$ in the definition of the trajectories that contribute to coherent backscattering, \textit{i.e.} we sum over trajectories that return from ${\bf A}^{}$ to ${\bf A}^{}$. The thereby introduced error is accounted for by expanding the action in the exponential up to first order in ${\bf A}^{\prime}-{\bf A}^{}$:
\begin{eqnarray}
\left(\tilde{R}_{\gamma}-\tilde{R}_{\gamma_p}\right)_{\rm cbs}&\simeq&2\hbar\left(\tilde{{\bf B}}^{(\gamma)}(t)+\tilde{{\bf B}}^{(\gamma)}(0)\right)\cdot\left(\tilde{{\bf A}}^{\prime}-\tilde{{\bf A}}^{}\right) \nonumber \\
&&+\text{terms independent of }\tilde{{\bf A}}^{\prime}\,,
\end{eqnarray}
where the subscript `$\rm cbs$' indicates that the action difference is evaluated choosing finally $\gamma_p=\mathcal{T}\gamma$ with $\mathcal{T}$ denoting (classical) time reversal. Here it has been used that $\tilde{{\bf B}}^{(\mathcal{T}\gamma)}(t)=-\tilde{{\bf B}}^{(\gamma)}(0)$. Performing now the integration over $\tilde{{\bf A}}^{\prime}$ yields
\begin{eqnarray}
&&\int{\rm d}^{L+1}\tilde{A}^{\prime}\exp\left[2{i}\left(\tilde{{\bf B}}^{(\gamma)}(t)+\tilde{{\bf B}}^{(\gamma)}(0)\right)\cdot\left(\tilde{{\bf A}}^{\prime}-\tilde{{\bf A}}^{}\right)\right]= \nonumber \\
&&\qquad \pi^{L+1}\delta\left(\tilde{{\bf B}}^{(\gamma)}(t)+\tilde{{\bf B}}^{(\gamma)}(0)\right)\,,
\end{eqnarray}
which is nonvanishing only if $\tilde{{\bf B}}^{(\gamma)}(t)=-\tilde{{\bf B}}^{(\gamma)}(0)=\tilde{{\bf B}}^{(\mathcal{T}\gamma)}(t)$, \textit{i.e.}, if the trajectory is self-retracing. In that case, however, we have $\gamma \equiv \mathcal{T}\gamma$ by construction. Pairing $\gamma$ with $\gamma_p=\mathcal{T}\gamma$ then amounts to pairing $\gamma$ with itself, which is already done on the level of the diagonal approximation. Hence, there are no coherent backscattering contributions within open systems, due to the non-conserved number of particles.


\begin{thebibliography}{10}
\bibitem{NoteFock}
This limitation mainly arises from the fact that the semiclassical
  approximation for the propagator in the Fock state representation involves a
  Wentzel-Kramers-Brillouin (WKB) approximation of the harmonic-oscillator
  eigenfunctions appearing in Eq.~\eqref{eq:quad2n}, which requires
  $n_\alpha\gg 1$ to be valid.
\bibitem{NoteErgodicity}
A technical complication as compared to closed systems emerges from the
  classical dynamics not necessarily being ergodic within a scattering system,
  due to the fact that atoms can be transferred to the leads before exploring
  the entire phase space within the scattering region. While this complication
  inhibits the direct application of ergodicity arguments in the evaluation of
  loop contributions, it does not affect the general conclusion that loop
  corrections vanish in the scattering context.
\end{thebibliography}
\end{document}